\documentstyle[preprint,aps,prd,psfig,here]{revtex}
\tightenlines
\newcommand{\be}{\begin{equation}}
\newcommand{\ee}{\end{equation}}
\newcommand{\bea}{\begin{eqnarray}}
\newcommand{\eea}{\end{eqnarray}}
\newcommand{\bean}{\begin{eqnarray*}}
\newcommand{\eean}{\end{eqnarray*}}
\newcommand{\ba}{\begin{array}}
\newcommand{\ea}{\end{array}}

\newcommand{\slashs}[1]{\not{\!#1}}
\def\Li{{\rm Li}_2}
\def\J#1{J_{#1}}
\def\N#1{N_{#1}}
\def\L#1{L_{#1}}
\def\K#1{K_{#1}}
\def\uint{\int dy\,dz\:\:}
\def\pint{\int{dy\,dz\over\sqrt{(1-y)^2-a}}\:\:}
\def\tint{\int{dy\,dz \over (1-y)^2-a}\:\:}
\def\kint{\int{dy\,dz\over\{(1-y)^2-a\}^{3/2}}\:\:}
\begin{document}
%-------------------- Title page -------------------------------
\draft
%-------------------- preprint N.O.----------------------------- 
\preprint{\vbox{\baselineskip=12pt
\rightline{HUPD-9818 }
\rightline{Fermilab-Pub-98/193-T}
\rightline{hep-ph/9807209}
          }     }
%-------------------- Title  -----------------------------------
\title{QCD Corrections to Spin Correlations\\ in Top Quark Production
         at Lepton Colliders}

\author{ Jiro KODAIRA, Takashi NASUNO }
\address{ Dept. of Physics, Hiroshima University\\
          Higashi-Hiroshima 739-8526, JAPAN }

\author{ Stephen PARKE }
\address{ Theoretical Physics Department \\
          Fermi National Accelerator Laboratory\\
          P. O. Box 500, Batavia, IL 60510, USA\\ }
\date{ \today }
\maketitle

\vspace{2cm}

%----------------------- abstract -------------------------------
\begin{abstract}
Spin correlations, using a generic spin basis, are investigated
to leading order in QCD for top quark production at lepton colliders. 
Even though, these radiative corrections
induce an anomalous $\gamma/Z$ magnetic moment for the top quarks
and allow for single, real gluon emission,
their effects on the top quark spin orientation are very small.
The final results are that  
the top (or anti-top) quarks are produced
in an essentially unique spin configuration in polarized lepton
collisions even after including the ${\cal O}( \alpha_{s} )$
QCD corrections.
\end{abstract}
%-------------------------- PACS N.O. -----------------------------
\vspace{2cm}


%
%----------------------- Text -----------------------------------
\section{Introduction}

The discovery of the top quark, with a mass near 175 GeV~\cite{CDF,D0}, 
provides us with a unique opportunity to better
understand electro-weak symmetry breaking and to search
for hints of physics beyond the standard model.
It has been known for sometime that top quarks decay electroweakly
before hadronization~\cite{k,bdkkz} and that
there are significant angular correlations between
the decay products of the top quark and the spin of the top quark~\cite{Decays}.
Therefore if the production mechanism of the top quarks correlates
the spins of the top and anti-top quarks, there will be a sizable angular
correlations between all the particles, both incoming and outgoing,
in these events.

There are many papers on the angular correlations for top
quark events~\cite{bop,kly} produced both at $e^+ e^-$
colliders~\cite{eecollider}
and hadron colliders~\cite{hcollider,mp}.
In most of these works, the top quark spin is decomposed in the
helicity basis. Recently, Mahlon and Parke~\cite{mp} have proposed
a more optimal decomposition of the top quark spin which results
in a large asymmetry at hadron colliders. Parke and Shadmi~\cite{ps}
extended this study to $e^+ e^-$ annihilation process at the
leading order in the perturbation theory and
found that the \lq\lq off-diagonal\rq\rq\ basis is the most
efficient decomposition of the top (anti-top) quark spin.
In this spin basis the top quarks are produced in an essentially unique spin 
configuration.
Since this result is of great interest, it is 
of crucial importance to estimate the
radiative corrections to this process which are dominated by QCD
effects.

The QCD corrections to top quark production
can be calculated perturbatively at energies sufficiently above
the production threshold of the top quark pairs.
The analytical study of QCD radiative corrections to heavy
quark production was pioneered in Ref.~\cite{gnt}
(see {\it e.g.} Ref.~\cite{gkl} for a recent article).
Polarized heavy quark production, in the helicity basis, 
has also been investigated by many authors~\cite{s,tung}.

In this article, we present an analytic differential cross section
for polarized top quark production at the QCD one-loop level.
We focus on the issue of what is the optimal decomposition of the
top quark spin for  $e^+ e^-$ colliders \footnote{
The physics of top quark production at muon colliders and $e^+e^-$ colliders 
is identical provided the energy is not tuned to the Higgs boson
resonance.}.
We have calculated
the cross section in a \lq\lq generic\rq\rq\ spin basis which
includes the helicity basis as a special case.
The radiative corrections, in general, add two effects to the tree
level analysis: the first is that a new vertex structure (anomalous
$\gamma/Z$ magnetic moment) is induced by the loop corrections to the
tree level
vertex, the second is that a (hard) real gluon emission from the final 
quarks can flip the spin and change the momentum of the parent quarks.
Therefore, compared to the radiative corrections to physical quantities which 
are spin independent,
it is possible that spin-dependent
quantities maybe particularly sensitive to the effects of QCD radiative
corrections.

The article is organized as follows. In Section 2, we examine the QCD
corrections to the polarized top (anti-top) quark production
in the soft gluon approximation. The aim of this section is:
(1) we estimate the numerical effects from the new vertex structure
on the spin correlation found in the tree level analysis and (2) we show  
that we can use the off-diagonal basis as a optimal basis also at the 
QCD one-loop level. In Section 3, we present our analytic calculations
of the full one-loop corrections to the polarized top quark production
in a generic spin basis. 
We give the numerical results
both in the helicity, beamline and the off-diagonal bases in Section 4. 
Here we
compare the full one loop results with those of the tree level and soft gluon
approximations. 
Finally, Section 5 contains the conclusions.
The phase space integrals which are needed in Section 4 are summarized
in Appendix A.
The unpolarized total cross section for top pair
production, using our results, is given in Appendix B as a cross check.

\section{Spin Correlations in the Soft Gluon Approximation}

In this section we derive the first order QCD corrected spin dependent,
differential cross section for top quark pair production 
in the soft gluon approximation(SGA).
It is instructive to first consider the soft gluon approximation
because in this approximation only the QCD vertex corrections
modify the spin correlations of the top quarks.
The full one-loop analysis will be given in the next section.
We use the same generic spin basis as in Ref.~\cite{ps}.
In this paper we do not consider transverse polarization of the top quarks
\footnote{It is known~\cite{kly} that the transverse top quark polarization
becomes nonzero when the higher order QCD corrections are included and
that this transverse polarization is very important and
related to the phenomena of CP violation.}
since we are interested in how QCD corrections modify the tree level spin
correlations and which spin basis is the most
effective for spin correlation studies.
Therefore we use a generic spin basis with the spin of the 
top quark and anti-top quark in the production plane.
We define the spins of the top and anti-top quarks by
the parameter $\xi$ as given in Fig.1.
%%%%%%%%%%%%%%%%%%%%%%%
% Fig.1
%%%%%%%%%%%%%%%%%%%%%%%
\noindent
The top quark spin is decomposed along the direction ${\bf s}_t$
in the rest frame of the top quark which makes an angle
$\xi$ with the anti-top quark momentum in the clockwise direction.   
Similarly, the anti-top quark spin states are defined in the anti-top
rest frame along the direction ${\bf s}_{\bar{t}}$ having the same
angle $\xi$ from the direction of the top quark momentum.
We use the following notation in this paper: the state
$t_{\uparrow}\,\bar{t}_{\uparrow}\,(t_{\downarrow}\,\bar{t}_{\downarrow})$
refers to a top with spin in the $+ {\bf s}_t \,(- {\bf s}_t )$
direction in the top rest frame and an anti-top
with spin $+ {\bf s}_{\bar{t}} \,(- {\bf s}_{\bar{t}} )$
in the anti-top rest frame.

The one-loop QCD correction to the cross section
is given by the interference between the tree and one-loop
vertex diagrams in Fig.2.
%%%%%%%%%%%%%%%%%%%%%%%
% Fig.2
%%%%%%%%%%%%%%%%%%%%%%%
\noindent
At the one-loop level, the $\gamma -t-\bar{t}$ and $Z-t-\bar{t}$ vertex
functions can be written in terms of three form factors
$A,B,C$ as follows:
\bea
   \Gamma^{\gamma}_{\mu} &=& e Q_t
        \left[ ( 1 + A ) \, \gamma_{\mu} + B \, \frac{t_{\mu} -
              \bar{t}_{\mu}}{2 m} \right] \ ,\label{gammavertex}\\
   \Gamma^Z_{\mu} &=& \frac{e}{\sin \theta_W}
        \biggl[ \left\{ Q_t^L \, ( 1 + A ) + ( Q_t^L - Q_t^R ) \, B \right\}
                (\gamma_L )_{\mu} \nonumber\\
         & & \qquad\qquad
          + \left\{ Q_t^R \, ( 1 + A ) -  ( Q_t^L - Q_t^R ) \, B \right\}
                (\gamma_R )_{\mu} \nonumber\\
          & & \qquad\qquad\qquad + \frac{Q_t^L + Q_t^R}{2} \, B\, 
           \frac{t_{\mu} - \bar{t}_{\mu}}{2 m} + \frac{Q_t^L - Q_t^R}{2} 
           \, C \, \frac{t_{\mu} + \bar{t}_{\mu}}{2 m} \gamma_5 \biggr]
              \ ,\label{zvertex}
\eea
where $Q_t = \frac{2}{3}$ is the electric charge of the top quark in units of
the electron charge $e$, $\theta_W$ is the Weinberg angle, and
$m$ and $t_{\mu}\, (\bar{t}_{\mu})$ are the mass and the momentum
of the top (anti-top) quark
($\gamma^{\mu}_{R/L} \equiv \gamma^{\mu} \frac{1 \pm \gamma_5}{2}$). 
The top quark couplings to the $Z$ boson are given by
\be
     Q_t^L = \frac{3 - 4 \sin^2 \theta_W}{6 \cos \theta_W} \quad ,\quad 
     Q_t^R = - \frac{2 \sin^2 \theta_W}{3 \cos \theta_W} \ .\label{zcoupling}
\ee
After multiplying the wave function renormalization factor (we employ
the on-shell renormalization scheme), the \lq\lq renormalized\rq\rq\ 
form factors read
\bea
    A &=& \hat{\alpha}_s     
     \left[ \left( \frac{1 + \beta^2}{\beta} \ln
                  \frac{1 + \beta}{1 - \beta} - 2 \right)
              \ln \frac{\lambda^2}{m^2} - 4 + 3 \beta
              \ln \frac{1 + \beta}{1 - \beta} \right.\nonumber\\
       &+& \left. \frac{1 + \beta^2}{\beta}
            \left\{ \frac{1}{2} \ln^2 \frac{1 + \beta}{1 - \beta}
           + \ln^2 \frac{1 + \beta}{2 \beta} -
                       \ln^2 \frac{1 - \beta}{2 \beta}
          + 2 {\rm Li}_2 \left( \frac{1 - \beta}{1 + \beta} \right)
          + \frac{2}{3} \pi^2 \right\} \right]\ ,\label{afactor}\\
    B &=& \hat{\alpha}_s
          \frac{1 - \beta^2}{\beta} \ln \frac{1 + \beta}{1 -
                      \beta}\ ,\label{bfactor}\\
    C &=& \hat{\alpha}_s
          \left[ ( 2 + \beta^2 ) \frac{1 - \beta^2}{\beta}
          \ln \frac{1 + \beta}{1 - \beta} -2 (1 - \beta^2 ) \right]\ ,
                \label{cfactor}
\eea
where $\beta$ is the speed of the produced top (anti-top)
quark and the strong coupling constant is $\hat{\alpha}_s \equiv
\frac{C_2 (R)}{4 \pi} \alpha_s = \frac{C_2 (R)}{(4 \pi )^2} g^2$
with $C_2 (R) = \frac{4}{3}$ for SU(3) of color.
We have introduced an infinitesimal mass $\lambda$ for the gluon to avoid
infrared singularities.
In the above expressions, we have shown only the real part of the
form factors because (1) the Z width is negligible in the region
of center-of-mass (CM) energy $\sqrt{s}$ far above the production
threshold for top quarks and (2) we are not considering the transverse
polarization for the top quarks. 
The contribution from $C$ can be neglected since it is proportional
to the electron mass.  

The differential cross section at the one loop level is given by
\bea
   \lefteqn{\frac{d \sigma}{d \cos \theta}
        \left( e^-_L e^+_R \to t_{\uparrow} \bar{t}_{\uparrow} \right)
        = \frac{d \sigma}{d \cos \theta}
        \left( e^-_L e^+_R \to t_{\downarrow} \bar{t}_{\downarrow} \right)}
          \nonumber \\
    &=& \left( \frac{3 \pi \alpha^2}{2 s} \beta \right)
          ( A_{LR} \cos \xi - B_{LR} \sin \xi ) \label{upup1loop}\\
    & & \qquad \times \Bigl[ \, ( A_{LR} \cos \xi - B_{LR} \sin \xi )
                 ( 1 + 2 A + 2 B)  \nonumber\\
    & & \qquad\qquad\qquad\qquad - \, 2\, 
          (\gamma^2 A_{LR} \cos \xi - \bar{B}_{LR} \sin \xi )\, B \, \Bigr]
               \nonumber \ ,\\
   \lefteqn{\frac{d \sigma}{d \cos \theta}
        \left( e^-_L e^+_R \to t_{\uparrow} \bar{t}_{\downarrow} 
                 \  {\rm or} \  t_{\downarrow} \bar{t}_{\uparrow} \right)} 
             \nonumber\\
    &=& \left( \frac{3 \pi \alpha^2}{2 s} \beta \right)
          ( A_{LR} \sin \xi + B_{LR} \cos \xi 
               \pm D_{LR} ) \label{updown1loop}\\
    & & \qquad \times \Bigl[ \, ( A_{LR} \sin \xi + B_{LR} \cos \xi 
           \pm D_{LR} ) ( 1 + 2 A + 2 B)  \nonumber\\
    & & \qquad\qquad\qquad\qquad  - \, 2\, 
          (\gamma^2 A_{LR} \sin \xi + \bar{B}_{LR} \cos \xi
              \pm \bar{D}_{LR} )\, B \, \Bigr]
               \nonumber .
\eea
Here, the angle $\theta$ is the scattering angle of the top quark with 
respect to the electron
in the zero momentum frame, $\alpha$ is the QED fine structure constant
and $\gamma = 1/\sqrt{1 - \beta^2}$. 
The quantities
$A_{LR},B_{LR},\bar{B}_{LR},D_{LR}$ and $\bar{D}_{LR}$ 
are defined by
\bea
    A_{LR} &=& \Bigl[ ( f_{LL} + f_{LR} ) \sqrt{1 - \beta^2}\,
            \sin \theta \Bigr] / \, 2 \ ,\nonumber\\
    B_{LR} &=& \Bigl[ f_{LL} ( \cos \theta + \beta )
             + f_{LR} ( \cos \theta - \beta ) \Bigr] / \, 2 = 
          \bar{B}_{LR} (- \beta ) \ ,\label{largecoupling}\\
    D_{LR} &=& \Bigl[ f_{LL} ( 1 + \beta \cos\theta )
             + f_{LR} ( 1 - \beta \cos\theta ) \Bigr] / \, 2 = 
          \bar{D}_{LR} (- \beta ) \ ,\nonumber
\eea
with
\[ f_{IJ} = - Q_t + Q_e^I Q_t^J \, \frac{1}{\sin^2 \theta_W}\,
              \frac{s}{s - M_Z^2} \ ,\]
where $M_Z$ is the $Z$ mass (as mentioned before, we neglect the $Z$
width) and $I,J \in (L,R)$. The electron couplings to the $Z$ boson
are
\[   Q_e^L = \frac{2 \sin^2 \theta_W - 1}{2 \cos \theta_W} \quad ,\quad 
     Q_e^R = \frac{ \sin^2 \theta_W}{ \cos \theta_W} \ .\]

The cross sections Eqs.(\ref{upup1loop},\ref{updown1loop}) 
contain an infrared singularity (in the form factor $A$) that
will be canceled  by the contributions from the real gluon
emission. In the soft gluon approximation, it is very easy to
calculate the real gluon contribution. As is well known, the amplitude
for the soft gluon emissions can be written in the factorized
form proportional to the tree amplitude. This means that
the soft gluon emission does not change the spin configurations 
or momentum of
the produced heavy quark pairs from the tree level values.
Therefore the QCD radiative corrections enter mainly through
the modifications of the vertex parts
Eqs.(\ref{gammavertex},\ref{zvertex}).
The cross section for the soft gluon emissions can be written as
\be
      \frac{d \sigma}{d \cos \theta} = J_{\rm IR}\ 
                     \frac{d \sigma_0}{d \cos \theta} \ ,\label{sgxsection}
\ee
where the subscript $0$ denotes the tree level
cross section. The soft gluon contribution $J_{\rm IR}$ is defined by
\[   J_{\rm IR} \equiv  - 4 \pi C_2 (R) \alpha_s
           \int^{k^0 =\omega_{\rm max}} \frac{d^3 \vec{k}}{(2 \pi )^3 2 k^0}
           \left( \frac{t_{\mu}}{t \cdot k} - \frac{\bar{t}_{\mu}}
           {\bar{t} \cdot k} \right)^2 \ ,\]
where $\omega_{\rm max}$ is the cut-off of the soft
gluon energy. This integral can be easily performed and we obtain
\bean
     J_{\rm IR} &=& 2 \hat{\alpha}_s  
              \left[  \left( \frac{1 + \beta^2}{\beta} \ln 
     \frac{1 + \beta}{1 - \beta} -2 \right) \ln \frac{4 \omega_{\rm max}^2}
       {\lambda^2} \right.\\
       & & \qquad\qquad + \left. \frac{2}{\beta} \ln \frac{1 + \beta}{1 - \beta}
       - \frac{1 + \beta^2}{\beta} \left[ 2 {\rm Li}_2 
              \left( \frac{2 \beta}{1 + \beta} \right)
            + \frac{1}{2} \ln^2 \frac{1 + \beta}{1 - \beta} \right] \right]\ .
\eean

By adding the one loop contributions
Eqs.(\ref{upup1loop},\ref{updown1loop})
and the soft gluon ones Eq.(\ref{sgxsection}), one can see that
the infrared singularities, $\ln \lambda$, are canceled out and
the finite results are obtained by replacing $2 A$ by
\bean
   \lefteqn{2 A + J_{\rm IR} 
}
\\       &=& 2 \hat{\alpha}_s     
     \left[ \left( \frac{1 + \beta^2}{\beta} \ln
                  \frac{1 + \beta}{1 - \beta} - 2 \right)
              \ln \frac{4 \omega_{\rm max}^2}{m^2} - 4 + 
           \frac{2 + 3 \beta^2}{\beta}
              \ln \frac{1 + \beta}{1 - \beta} \right.\\
       &+& \left. \frac{1 + \beta^2}{\beta}
            \left\{ \ln \frac{1 - \beta}{1 + \beta}
               \left( 3 \ln \frac{2 \beta}{1 +  \beta}
         +  \ln \frac{2 \beta}{1 - \beta} \right) 
         + 4 {\rm Li}_2 \left( \frac{1 - \beta}{1 + \beta} \right)
          + \frac{1}{3} \pi^2 \right\} \right]\ ,
\eean
in Eqs.(\ref{upup1loop},\ref{updown1loop}).

The cross sections for $e^-_R e^+_L$ can be obtained by interchanging
$L,R$ as well as $\uparrow , \downarrow$ in the above formulae.

Since we are interested in maximizing the spin correlations
of the top quark pairs we must vary the spin angle, $\xi$, 
to find the appropriate spin basis.
At tree level, it is known that there exists 
the \lq\lq off-diagonal\rq\rq\ basis 
which makes the contributions from the like-spin 
configuration vanish~\cite{ps}. 
At order ${\cal O}(\alpha_s )$, we 
find that definition of the off-diagonal basis for $e^-_Le^+_R$
scattering is not modified
by QCD corrections,  with the spin angle, $\xi$, 
satisfying the tree level relationship
\be
 \tan \xi = \frac{A_{LR}}{B_{LR}}
   = \frac{( f_{LL} + f_{LR} ) \sqrt{1 - \beta^2}\, \sin \theta}
          {f_{LL} ( \cos \theta + \beta )
             + f_{LR} ( \cos \theta - \beta )} \ .\label{offdiaxi}
\ee
The first order QCD corrected cross sections in this basis are
\bea
   \lefteqn{\frac{d \sigma}{d \cos \theta}
    \left( e^-_L e^+_R \to t_{\uparrow} \bar{t}_{\uparrow} 
           \ {\rm and} \ t_{\downarrow} \bar{t}_{\downarrow} \right)
     = 0 \ ,}\label{upupsoft}\\
   \lefteqn{\frac{d \sigma}{d \cos \theta}
        \left( e^-_L e^+_R \to t_{\uparrow} \bar{t}_{\downarrow} 
                 \  {\rm or} \  t_{\downarrow} \bar{t}_{\uparrow} \right)
     = \left( \frac{3 \pi \alpha^2}{2 s} \beta \right)
      \Bigl( \sqrt{A_{LR}^2 + B_{LR}^2} \mp D_{LR} \Bigr)} \nonumber \\
       &\times& \Biggl[ \Bigl( \sqrt{A_{LR}^2 + B_{LR}^2} \mp D_{LR} \Bigr)
        \Bigl( 1 + S_I \Bigr) 
      - 2 \, \left( \frac{\gamma^2 A_{LR}^2 + B_{LR} \bar{B}_{LR}}
            {\sqrt{A_{LR}^2 + B_{LR}^2}} \mp \bar{D}_{LR} \right)
                S_{II} \Biggr]  \ ,\label{updownsoft}
\eea
where
\[  S_I = 2 A + J_{\rm IR} + 2 B \quad , \quad S_{II} = B \ .\]
A similar result holds for $e^-_Re^+_L$ scattering. 

In Fig.3 we show the differential cross sections in the off-diagonal basis,
Eq.(\ref{offdiaxi}),
for  $\sqrt{s} = 400 \, {\rm GeV}$.
%%%%%%%%%%%%%%%%%%%%%%%%%%
% Fig.3
%%%%%%%%%%%%%%%%%%%%%%%%%%
\noindent
The following values for the parameters of the standard model were used
\[  m \equiv M_{\rm top} = 175 \, {\rm GeV} \quad , \quad
      M_Z = 91.187 \, {\rm GeV} \ ,\]
\[  \alpha = \frac{1}{128} \quad , \quad \alpha_s (M_Z^2 ) = 0.118
            \quad , \quad \sin^2 \theta_W = 0.2315 \ .\]
We use $\sqrt{s}$ as the renormalization scale and the pole mass for the 
top quark.
For $e^-_Le^+_R$ scattering the up-up $(t_{\uparrow} \bar{t}_{\uparrow})$ 
and the down-down
$t_{\downarrow} \bar{t}_{\downarrow}$ components are identically zero.
The total cross section is more than $99\%$ up-down
$t_{\uparrow} \bar{t}_{\downarrow}$ and less than $1\%$ down-up
$t_{\downarrow} \bar{t}_{\uparrow}$.
For $e^-_Re^+_L$ scattering the up-up $(t_{\uparrow} \bar{t}_{\uparrow})$ 
and the down-down
$t_{\downarrow} \bar{t}_{\downarrow}$ components are 
non-zero because we have used the off-diagonal basis for $e^-_Le^+_R$
scattering. However the down-up
$t_{\downarrow} \bar{t}_{\uparrow}$ component is still more than $99\%$
of the total cross section.

Although there exists a magnetic 
moment modification to the $\gamma/Z-t-\bar{t}$ vertex from
QCD corrections, this does not change the behavior of the 
spin dependent cross sections in the off-diagonal basis. 
The QCD corrections, however, make the
differential cross sections larger by $\sim 30 \%$ compared to the
tree level ones at this $\sqrt{s}$. 
Thus the off-diagonal basis continues to display very strong spin correlations 
for the top quark pairs even after taking 
the QCD corrections into account, at least in the soft gluon approximation.

In the previous paragraph the cut-off energy for the
soft gluon has been chosen as $\omega_{\rm max} = 10 \, {\rm GeV}$.
The results, of course, depend on the value of $\omega_{\rm max}$.
The $\omega_{\rm max}$ dependence of the cross section is examined in
Fig.4. 
The cross section behaves quite uniformly as the 
value of $\omega_{\rm max}$ is changed thus the above conclusions remain
qualitatively the same for any reasonable value of $\omega_{\rm max}$. 

%%%%%%%%%%%%%%%%%%%%%%%%%%%%
% Fig.4
%%%%%%%%%%%%%%%%%%%%%%%%%%%%
\noindent

\section{Single Spin Correlations in $ e^+ e^- $ Process}

The soft gluon approximation used in the last section has 
two shortcomings. 
First, soft gluon emission cannot change the spin 
of the heavy quarks and second the heavy quark pairs are always 
back to back.
Neither of these approximations is valid for hard gluon emission.
Hence it is possible that the full 
${\cal O}(\alpha_s )$ QCD corrections might 
completely change the conclusions of the previous section.
Therefore, in this section, we investigate the full ${\cal O}(\alpha_s )$
QCD corrections.
Since, in the presence of a hard gluon, 
the top and anti-top quarks are not generally produced back-to-back,
it is more sensible to consider the
single heavy quark spin correlations,
namely the inclusive cross section for the production
of the top (or anti-top) quark in a particular spin configuration.
We have organized this section as follows.
After defining the kinematics and our conventions, we give the polarized
cross section for the top quark using a generic spin basis closely
related to the spin basis of the previous section.
The numerical analysis will be relegated to the next section.

\subsection{Amplitudes and Kinematics}

The principle result of this section will be 
the inclusive cross section for polarized
top quark production, $e^+ e^- \to t_{\uparrow}{\rm ~or~}t_{\downarrow} ~+~ X$, 
where $X = \bar{t} {\rm ~or~} \bar{t} g$.
(The cross section for the anti-top quark inclusive production can be 
easily obtained from the results in this section.)
Since the vertex corrections are the same as in the previous
section all that is required is the full real gluon emission contributions.
The real gluon emission diagrams to leading order 
in $\alpha_s$ are given in Fig.5.
%%%%%%%%%%%%%%%%%%%%%%%%%%%%%%%%%
% Fig.5
%%%%%%%%%%%%%%%%%%%%%%%%%%%%%%%%%
This figure also defines the momenta of particles.
We will use the spinor helicity method for massive fermions~\cite{mp}
to calculate the squares of these amplitudes for a polarized top 
quark.
The top quark momentum $t$ is decomposed into a sum of two massless
momenta $t_1 \,,\,t_2$ such that in the rest frame of the top
quark the spatial momentum of $t_1$ defines the spin axis for the top
quark.
\[    t = t_1 + t_2 \quad , \quad m s_t = t_1 - t_2 \ ,\]
where $s_t$ is the spin four vector of the top quark.

The amplitude for Fig.5 is given by
\bea
 \lefteqn{M (e_L^- e_R^+ \to t_{\alpha} \bar{t}_{\beta} g)
   = \bar{v}_{\uparrow} (\bar{q}) \gamma_L^{\mu} u_{\downarrow} (q)}
                \nonumber\\
   &\times& \bar{u}_{\alpha} (t) \left[ \frac{1}{2 \bar{t} \cdot k}
             \left( \frac{a_{LL}}{2} \gamma_L^{\mu} +
                    \frac{a_{LR}}{2} \gamma_R^{\mu} \right)
               ( - \bar{\slashs{t}} - \slashs{k} + m ) \gamma_{\nu} \right.
                 \label{amplitude}\\
   & & \qquad\qquad + \left. \frac{1}{2 t \cdot k} \gamma_{\nu}
               ( \slashs{t} + \slashs{k} + m )
             \left( \frac{a_{LL}}{2} \gamma_L^{\mu} +
                    \frac{a_{LR}}{2} \gamma_R^{\mu} \right)
             \right] T^a v_{\beta} (\bar{t}) \varepsilon^{\nu}_a (k)
                \ ,\nonumber
\eea
where $\alpha , \beta$ are the spin indices for quarks
and $\varepsilon$ is the
polarization vector of the gluon. $T^a$ is the color matrix.
The coupling constants $a_{LI}$ are defined as follows:
\[  \frac{a_{LI}}{2} = \frac{e^2 g}{s} f_{LI} \ .\]

The expressions for the squares of the amplitudes given below
have been summed over the spins of the unobserved particles
(the anti-top quark and gluon) as well as the colors of the final
state particles. 
Let us write the square of the amplitude for the top quark
with spin \lq\lq up\rq\rq\ as
\bea
  \lefteqn{|M (e_L^- e_R^+ \to t_{\uparrow} \bar{t} g )|^2}\nonumber\\ 
    &=& N_c C_2 (R) \, \left[ \frac{1}{(\bar{t} \cdot k )^2}\, T_1 +
        \frac{1}{(\bar{t}\cdot k)(t\cdot k)}\, T_2 +
         \frac{1}{(t \cdot k )^2}\, T_3 \right] \ .\label{ampsquare}
\eea
After some calculation, we find
\bean
   T_1 &=& 4 \, |a_{LL}|^2 
        \left[ (\bar{t}\cdot k)(q\cdot k) - m^2 q\cdot (\bar{t} + k) 
          \right] \, (t_2 \cdot \bar{q}) \\
        &+& \, 4 \, |a_{LR}|^2
        \left[ (\bar{t}\cdot k)(\bar{q}\cdot k) - m^2 
            \bar{q} \cdot (\bar{t} + k) 
          \right] \, (t_1 \cdot q) \\
        &-&  \,  \left[ a_{LL} a_{LR}^* 
           ( (\bar{t}\cdot k) + m^2 ) {\rm Tr} (\omega_+ \, t_1 \, t_2 \,
            \bar{q} \, q ) + c.c. \right] \ ,\\
   T_2 &=& |a_{LL}|^2 
         \left[ 4 (t\cdot\bar{t})(\bar{t}\cdot q)(t_2 \cdot\bar{q})
         + (t_2 \cdot\bar{q}) {\rm Tr} (\omega_+ \, q \, k \, t \, \bar{t})
         + (\bar{t}\cdot q) {\rm Tr} (\omega_- \, t_2 \, \bar{q} \, k \, 
       \bar{t}) \right] \\
        &+& \, |a_{LR}|^2 
         \left[ 4 \, (t\cdot\bar{t})(\bar{t}\cdot\bar{q})(t_1 \cdot q)
         + (t_1 \cdot q) {\rm Tr} (\omega_- \, \bar{q} \, k \, t \, \bar{t})
         + (\bar{t}\cdot\bar{q}) {\rm Tr} (\omega_+ \, t_1 \, q \, k \, 
           \bar{t}) \right] \\
        &+& \, a_{LL} a_{LR}^* 
           \Bigl[ (t\cdot\bar{t})
                 {\rm Tr} (\omega_+ \, t_1 \, t_2 \, \bar{q} \, q)
          - (k\cdot q) {\rm Tr} (\omega_+ \, t_1 \, t_2 \, \bar{q} \, k)\\
        & & \qquad\qquad\qquad\quad
               + \frac{1}{2} {\rm Tr}
                 (\omega_+ \, t_2 \, t_1 \, q \, k \, t \, \bar{q}) 
               + \frac{1}{2} {\rm Tr}
                (\omega_+ \, t_1 \, t_2 \, \bar{q} \, q \, k \, \bar{t} ) 
              \Bigr] \\
        &+& \, a_{LL}^* a_{LR} 
           \Bigl[ (t\cdot\bar{t}) {\rm Tr}
                (\omega_- \, t_2 \, t_1 \, q \, \bar{q})
          - (k\cdot\bar{q}) {\rm Tr} (\omega_- \, t_2 \, t_1 \, q \, k)\\
        & & \qquad\qquad\qquad\quad
               + \frac{1}{2} {\rm Tr} 
           (\omega_- \, t_1 \, t_2 \, \bar{q} \, k \, t \, q ) 
               + \frac{1}{2} {\rm Tr}
                (\omega_- \, t_2 \, t_1 \, q \, \bar{q} \, k \, \bar{t} )
                \Bigr] \\
        &+& \, \, \, c.c. \ , \\
   T_3 &=& 2 \, |a_{LL}|^2
        \left[ - m^2 (k\cdot\bar{q}) - 2 (m^2 + (t\cdot k))
            (t_2 \cdot\bar{q}) + 2 ( (t + k)\cdot\bar{q})
                   (t_2 \cdot k) \right] \, (\bar{t}\cdot q )\\
       &+&  \, 2 \, |a_{LR}|^2
        \left[ - m^2 (k\cdot q) - 2 (m^2 + (t\cdot k))
            (t_1 \cdot q) + 2 ( (t + k)\cdot q) 
           (t_1 \cdot k )\right] \, (\bar{t}\cdot\bar{q}) \\
       &-&  \, \frac{m^2}{2}
           \left[ a_{LL} a_{LR}^* \{ 2 {\rm Tr}
                 (\omega_+ \, t_1 \, t_2 \, \bar{q} \, 
               q ) + {\rm Tr} (\omega_+ \, k \, t_2 \, \bar{q} \, q )  
                 + {\rm Tr} (\omega_+ \, t_1\, k \, \bar{q} \, q ) \} +
               c.c. \right] \ , 
\eean
where $\omega_{\pm} \equiv \frac{1 \pm \gamma_5}{2}$ and
all momentum, $p$, under the \lq\lq Tr\rq\rq\  operator are understood  
to be  $\slashs{p}$. By interchanging the $t_1$ and
$t_2$ vectors in the above expressions,
we can get the amplitude square for the top quark
with spin \lq\lq down\rq\rq.
Since we neglect the $Z$ width in this paper, all the coupling
constants $a_{LI}$ are real.

To define the spin basis for the top quark we naturally extend the spin 
definition of the previous section to the present case. The top quark
spin is decomposed along the direction ${\bf s}_t$ in the rest frame of the
top quark which makes an angle
$\xi$ with the sum of the anti-top quark and the gluon momenta
in the clockwise direction, see Fig.6.

%%%%%%%%%%%%%%%%%%%%%%%%%%%%%
% Fig.6
%%%%%%%%%%%%%%%%%%%%%%%%%%%%%

To calculate the cross section from Eq.(\ref{ampsquare}), we take the
CM frame in which the $e^+ e^-$ beam line coincides with the $z$-axis,
\[  q = \frac{\sqrt{s}}{2} ( 1\,,\,0\,,\,0\,,\,1 )\quad , \quad
    \bar{q} = \frac{\sqrt{s}}{2} ( 1\,,\,0\,,\,0\,,\,-1 )\ .\]
We specify the variables $x, y$ and $z$ which are related to CM 
energies of the gluon, top and anti-top quarks by  
\[  x \equiv 1 - \frac{2\, k\cdot (q + \bar{q})}{s}\ ,\ 
    y \equiv 1 - \frac{2\, t\cdot (q + \bar{q})}{s}\ ,\ 
    z \equiv 1 - \frac{2\, \bar{t}\cdot (q + \bar{q})}{s}\ .\] 
The momenta of the final state particles, in terms of these variables, are
\[        k = \frac{\sqrt{s}}{2} ( 1 - x \,,\, (1 - x) \, \hat{k} )\ , \,
t = \frac{\sqrt{s}}{2} ( 1 - y \,,\, a(y) \, \hat{t} )\ ,\     
     \bar{t} = \frac{\sqrt{s}}{2} ( 1 - z \,,\, a(z) \, \hat{\bar{t}} \, )\ \ 
,\]
where \ $\hat{}$ \ means the unit space vector and
\[  a(y) \equiv \sqrt{(1-y)^2 - a} \quad , \quad
        a(z) \equiv \sqrt{(1-z)^2 - a} \ ,\]
with $a \equiv 4 m^2 / s$. 
Fig.7 defines the orientation of the top and anti-top momenta
and by energy-momentum conservation the momentum of the
gluon is also determined.\footnote{
Our Fig.7 contains an extra degree of freedom, $\phi$,
compared to Fig.3 of the first paper in ref.~\cite{tung}. 
These works of Tung et al. correspond to setting our $\phi=\pi/2$.
Due to this difference we have been unable to reproduce their results. 
Our variable $\phi$ corresponds to the 
variable $\phi_{12}$ in the first paper of 
ref.~\cite{gnt}. }
%%%%%%%%%%%%%%%%%%%%%%%%%%%%%%%%
% Fig.7
%%%%%%%%%%%%%%%%%%%%%%%%%%%%%%%%
One can easily obtain the spin four vector
of the top quark in the CM frame by boosting the spin vector characterized
by $\xi$ in the top quark rest frame in the direction of top quark
momentum by $\beta (y)$ (the speed of the top quark in the CM frame).
The explicit form for $t_1 \, (t_2 = t - t_1 )$ is given by
\bea
      t_1^0 &=& \frac{m}{2} \left[ \, \gamma (y) \, (1 - \beta (y) \cos \xi )\,
               \right] \ ,\nonumber\\
      t_1^1 &=& \frac{m}{2} \left[ \,  \gamma (y) \, (\beta (y) - \cos \xi )
                  \sin \theta \cos \varphi + \sin \xi \cos \theta
                  \cos \varphi \, \right] \ ,\label{spinvecCM}\\
      t_1^2 &=& \frac{m}{2} \left[ \,  \gamma (y) \, (\beta (y) - \cos \xi )
                  \sin \theta \sin \varphi + \sin \xi \cos \theta
                  \sin \varphi \, \right] \ ,\nonumber\\
      t_1^3 &=& \frac{m}{2} \left[ \,  \gamma (y) \, (\beta (y) - \cos \xi )
                  \cos \theta - \sin \xi \sin \theta \, \right] \ , \nonumber
\eea
where
\[ \sqrt{a} \, \gamma (y) = 1 - y \quad , 
                    \quad \sqrt{a} \, \gamma (y) \, \beta (y) = a (y) \ .\]

If we eliminate the gluon momentum $k$ using the energy momentum
conservation and use the angular variables $\chi \,,\, \phi$ in
Fig.7 to specify the orientation of the anti-top quark
(if one eliminates the anti-top
momentum, one can proceed in the similar way by introducing
other angular variables), the square of the amplitude
Eq.(\ref{ampsquare}) can be written as
\be
  |M (e_L^- e_R^+ \to t_{\uparrow} \bar{t} g )|^2
     = \frac{s}{8}\, N_c C_2 (R) \left[ a_{LL}^2 M_1
           + a_{LR}^2 M_2 + a_{LL} a_{LR} M_3 \right] \ ,\label{ampsquareyz}
\ee 
where $M_i$ are the functions of $y\,,\,z$, angles defined above and
the spin orientation $\xi$.
\bean
   M_1 &=& \frac{2}{yz} \left[ (1-y)^2 + a^2 (y) \cos^2 \theta + (1-z)^2
                   + a^2 (z) \cos^2 \bar{\theta} \right]\\
       &-& \, a \left( \frac{1}{yz} + \frac{1}{y^2} \right)
             (1 - y^2 + a^2 (y) \cos^2 \theta )
              - a \left( \frac{1}{yz} + \frac{1}{z^2} \right)
             (1 - z^2 + a^2 (z) \cos^2 \bar{\theta} )\\
       &+& \, 2 \left( \frac{2-2y-a}{yz} - \frac{a}{y^2} \right)
             a (y) \cos \theta -
              2 \left( \frac{2-2z-a}{yz} - \frac{a}{z^2} \right)
             a (z) \cos \bar{\theta}\\
       &+& \, \frac{1}{y z^2} (1 - z - a(z) \cos \bar{\theta})
               \left[ y (1 - a(z) \cos \bar{\theta}) - 
                  z (1 - a(y) \cos \theta ) \right] (\delta t \cdot \bar{t}) \\
       &-& \, \frac{1}{y z^2} (1 - z - a(z) \cos \bar{\theta})
               \left[ y - z + (y+z)( z - a(z) \cos \bar{\theta}) \right]
                  (\delta t \cdot ( \bar{q} + q  )) \\
       &+& \, \frac{a}{yz} \left( \frac{1}{y} + \frac{1}{z} \right) 
               \left[ y (1 - a(z) \cos \bar{\theta}) + 
                  z (1 + a(y) \cos \theta ) \right]
                  (\delta t \cdot \bar{q}) \\
       &-& \, \frac{2}{yz} \left[ (1 - y + a(y) \cos \theta ) + 
                   (1 - y - z) (1 - z - a(z) \cos \bar{\theta}) \right]
                   (\delta t \cdot \bar{q}) \ , \\
   M_2 &=&  M_1 ( \cos \theta \to - \cos \theta \,,\,
                    \cos \bar{\theta} \to - \cos \bar{\theta} \,,\,
                    \delta t \to - \delta t ) \ , \\
   M_3 &=&  2a \left\{ \frac{4}{yz} - \frac{4}{y} - \frac{4}{z}
             - \frac{(y+z)^2}{yz} + \frac{(y+z)^2}{yz} \cos^2 \theta_k
             \right\} - 2 a^2 \left( \frac{1}{y} + \frac{1}{z} \right)^2\\
       &+& \, \frac{a}{y z} \left( \frac{1}{y} + \frac{1}{z} \right)
                 (y a(z) \cos \bar{\theta} - z a(y) \cos \theta)
                (\delta t \cdot ( \bar{q} + q )) \\ 
       &-& \, \frac{2}{y z}
              \left[ a(y) \cos \theta + (1 - y - z) a(z) \cos
               \bar{\theta} \right]
                (\delta t \cdot (\bar{q} + q )) \\ 
       &+& \, a \left( \frac{1}{y} + \frac{1}{z} \right)^2
                (\delta t \cdot (\bar{q} - q )) 
           +  \frac{2}{y z} \left[ z (y + z) - 2 (1 - y - z) \right]
                (\delta t \cdot (\bar{q} - q )) \\ 
       &+& \, \frac{2}{y z} \left[ (1 - y) a(z) \cos \bar{\theta} 
               + (3 + z) a(y) \cos \theta \right] (\delta t \cdot \bar{t}) \ .
\eean
In the above equations, $\theta_k$ is the angle between the $z$ axis
and the gluon momentum
\[  (y+z) \cos \theta_k = - a (y) \cos \theta - a (z) \cos \bar{\theta} \ ,\]
and $\delta t$ is defined as
\[  \delta t \equiv \frac{4}{s} ( t_1 - t_2 )\ .\]
Using Eq.(\ref{spinvecCM})
we find that the products of $\delta t$ with 
momenta $q\,,\,\bar{q}$ and $\bar{t}$
are 
\bean
   \delta t \cdot q &=& \{ (1-y) \cos \theta - a (y) \} \cos \xi
             + \sqrt{a} \sin \theta \sin \xi \ ,\\
   \delta t \cdot \bar{q} &=& \{ - (1-y) \cos \theta - a (y) \} \cos \xi
             - \sqrt{a} \sin \theta \sin \xi \ ,\\
   \delta t \cdot \bar{t} &=& \{ - (1-z) a (y) + (1-y) a (z) \cos \chi \}
                    \cos \xi\\
      & & \qquad\qquad\qquad\qquad\qquad
                    +\, \sqrt{a} a (z) \sin \xi \sin \chi \cos \phi \ .
\eean
The unpolarized top quark production process is given by dropping the spin
dependent parts (terms proportional to $\delta t$)
in Eq.(\ref{ampsquareyz}). 
As a check we have reproduced the results of Ref.~\cite{gnt} by putting 
$a_{LL} = a_{LR} = - \frac{2 e^2 g}{s} Q_q$.

The cross section is given by
\be
    d \sigma (e_L^- e_R^+ \to t_{\uparrow} \bar{t} g ) = 
    \frac{1}{2s} |M (e_L^- e_R^+ \to t_{\uparrow} \bar{t} g )|^2
             (PS)_3 \ , \label{diffcrosssection}
\ee
where $(PS)_3$ is the three particle phase space.
\[  (PS)_3 = \frac{d^3 t}{(2\pi )^3 2 t^0}
             \frac{d^3 \bar{t}}{(2\pi )^3 2 \bar{t}^0}
             \frac{d^3 k}{(2\pi )^3 2 k^0} (2\pi )^4 \delta^4
             (t + \bar{t} + k - q - \bar{q} ) \ .\]
We introduce a small mass $\lambda$
for the gluon to regularize the infrared singularities.
It is easy to rewrite the above phase space integral as,
\bean
 \int (PS)_3 &=&  \frac{s}{(4 \pi)^5}
         \int^{y_+}_{y_-} dy \int^{z_+ (y)}_{z_- (y)} dz\\ 
      & & \qquad\qquad\qquad \times \int d\Omega d\cos \chi d\phi \, \delta
      \left( \cos \chi - \frac{y + z + yz + a - 1 - 2 a_{\lambda}}
                  {a(y) a(z)} \right) \ ,
\eean
where $d\Omega = d\cos \theta d\varphi$ is the solid angle for the top
quark and $a_{\lambda} \equiv \lambda^2 / s$.
The integration regions over $y$ and $z$ are determined by the
condition $|\cos \chi | \leq 1$,
\bean
     y_+ &=& 1 - \sqrt{a} \quad , \quad y_- = \sqrt{a a_{\lambda}}
              + a_{\lambda} \ ,\\
     z_{\pm}(y) &=& \frac{2}{4 y + a}
           \left[ y \left( 1 - y - \frac{a}{2} + a_{\lambda} \right)
          + a_{\lambda} \pm
           a (y) \sqrt{( y - a_{\lambda})^2 - a a_{\lambda}} \right] \ .
\eean
The integration over the angle $\phi$ is not difficult if one uses
the relation
\[ \cos \bar{\theta} = \cos \theta \cos \chi
          + \sin \theta \sin \chi \cos \phi \ .\]
The integrals we need are the following:
\bean
  \int \cos\bar{\theta} \ d \phi &=& 2 \pi \cos\theta \cos\chi \ ,\\
  \int \cos^2\bar{\theta} \ d \phi &=& 2 \pi \left[ \cos^2\theta
      + \frac{1}{2} ( 1 - 3 \cos^2\theta ) \sin^2\chi \right] \ ,\\
  \int \cos^2\theta_k \ d \phi &=& 2 \pi \left[ \cos^2\theta
      + \frac{1}{2} ( 1 - 3 \cos^2\theta ) \ \frac{a^2 (z)}{(y + z)^2}
                       \  \sin^2\chi \right] \ ,\\
  \int \cos\bar{\theta} \cos\phi \ d \phi &=& \pi \sin\theta \sin\chi \ ,\\
  \int \cos^2\bar{\theta} \cos\phi \ d \phi &=&
             2 \pi \sin\theta \cos\theta \sin\chi \cos\chi \ .
\eean
Due to the $\delta$ function in the phase space integral,
the angle $\chi$ is a function of $y$ and $z$:
\bean
    \cos\chi &=& \frac{y + z + yz + a - 1}{a(y) a(z)}\ ,\\
    \sin^2\chi &=& \frac{4yz (1-z-y)- a (y+z)^2}{a^2 (y) a^2 (z)}\ ,
\eean
where we have put $a_{\lambda}$ to be zero since the $\lambda \to 0$
limit does not produce any singularities in the (squared) amplitude.
The remaining integrals to get the cross section are over the
variables $y$ and $z$. According to the type of
integrand, we group the phase space integrals (after the integrations
over the angular variables) into
four distinct classes $\{J_i\}\,,\,\{N_i\}\,,\,\{L_i\}$ and $\{K_i\}$~\cite{tung}.
The individual integrals of these classes are summarized in Appendix A.

\subsection{Cross Section in the Generic Spin Basis}

We write the inclusive cross section for the top quark
in the following form.
\be
 \frac{d \sigma}{d \cos \theta}
                  ( e_L^-  e_R^+ \to t_{\uparrow} X )
  = \frac{3 \pi \alpha^2}{4 s}
        \sum_{klmn} \left( D_{klmn} + \hat{\alpha}_s C_{klmn} \right)
              \cos^k \theta \sin^l \theta
             \cos^m \xi \sin^n \xi \ ,\label{txsection}
\ee
where $D_{klmn}$ are the contributions from the tree and the one-loop
diagrams and $C_{klmn}$ are from the real emission diagrams.
Let us first write down the $D_{klmn}$,
\bean
  D_{0000} &=& \beta [ f_{LL}^2 + f_{LR}^2 + 2 a f_{LL} f_{LR} ]
                    ( 1 + \hat{\alpha}_s V_{I} )\\
           & & \qquad\qquad\qquad - \beta [ 2 (f_{LL} + f_{LR})^2 - \beta^2
                (f_{LL} - f_{LR})^2 ] \hat{\alpha}_s V_{II} \ ,\\
  D_{2000} &=& \beta^3 (f_{LL}^2 + f_{LR}^2 )
                    ( 1 + \hat{\alpha}_s V_{I} ) + \beta^3
              (f_{LL} - f_{LR})^2 \hat{\alpha}_s V_{II} \ ,\\
  D_{1000} &=& 2 \beta^2 (f_{LL}^2 - f_{LR}^2 )
                    ( 1 + \hat{\alpha}_s V_{I} ) \ ,\\
  D_{0010} &=& \beta^2 (f_{LL}^2 - f_{LR}^2 )
                    ( 1 + \hat{\alpha}_s V_{I} ) \ ,\\
  D_{2010} &=& \beta^2 (f_{LL}^2 - f_{LR}^2 )
                    ( 1 + \hat{\alpha}_s V_{I} ) \ ,\\
  D_{1010} &=& \beta [ (f_{LL} + f_{LR} )^2 + 
              \beta^2 (f_{LL} - f_{LR} )^2  ]
                    ( 1 + \hat{\alpha}_s V_{I} )\\
           & & \qquad\qquad\qquad - 2 \beta [ (f_{LL} + f_{LR})^2 - \beta^2
                (f_{LL} - f_{LR})^2 ] \hat{\alpha}_s V_{II} \ ,\\
  D_{0101} &=& \frac{\beta}{\sqrt{a}} (f_{LL} + f_{LR} )^2 
                 [ a ( 1 + \hat{\alpha}_s V_{I} ) - ( 1 + a)
                       \hat{\alpha}_s V_{II} ] \ ,\\
  D_{1101} &=& \frac{\beta^2}{\sqrt{a}} (f_{LL}^2 - f_{LR}^2 ) 
                 [ a ( 1 + \hat{\alpha}_s V_{I} ) - ( 1 - a)
                       \hat{\alpha}_s V_{II}]\ ,
\eean
with
\[  \beta = \beta (0) = \sqrt{1-a} \quad , \quad
    \hat{\alpha}_s V_{I} = 2 A + 2 B \quad , \quad 
    \hat{\alpha}_s V_{II} = B \ .\]
$A\,,\,B$ are defined in Eqs.(\ref{afactor},\ref{bfactor}).
 
For the $C_{klmn}$ we find,
\bean
  C_{0000} &=& 2\  (f_{LL}^2 + f_{LR}^2 )\  \left[ \ J_{\rm IR}^1 
                  - (4 - a) J_3 + (2 + a) J_2 + a J_1 
               + \frac{1}{4} R_1 \right]\\
           & & + \  4 a \,f_{LL} f_{LR} \ \left[ \ J_{\rm IR}^1 
               - 4 J_3 - J_2 - J_1 + \frac{1}{4} R_2 \right] \ ,\\
  C_{2000} &=& 2\  (f_{LL}^2 + f_{LR}^2 ) \ \left[ \ \ (1 - a) J_{\rm IR}^1
               - (4 - a) J_3 + (2 - a) J_2 - a J_1  - \frac{3}{4} R_1 \right] \\
           & & + \  4 a \,f_{LL} f_{LR} \ \left[ \  J_2 +  J_1
                   - \frac{3}{4} R_2 \right] \ ,\\
  C_{1000} &=& 2\ (f_{LL}^2 - f_{LR}^2 ) \ \Bigl[ \ (1 - a) J_{\rm IR}^2
             + \ a N_{10} - 2 (4 - 3a) N_9 \\
           & & \qquad\qquad\qquad\qquad\quad  - \, (4 - 5a) N_8
                     - 2 N_7  + 2 N_6 + 6 N_3 + 2 N_2 \Bigr] \ ,\\
  C_{0010} &=& \frac{1}{2} \ C_{1000} \\ 
           & & + \  (f_{LL}^2 - f_{LR}^2 ) \ \left[ \  - 4 N_6
               - a N_4 - a N_3 - a N_2 + (4 - a) N_1 + 
                  \frac{1}{2} R_3 \right] \ ,\\
  C_{2010} &=& C_{0010} - 2 (f_{LL}^2 - f_{LR}^2 ) R_3 \ ,\\
  C_{1010} &=& (f_{LL}^2 + f_{LR}^2 ) \ \Bigl[ \ 2 (2 - a) J_{\rm IR}^1
               + 2 a J_4 - 2 (8 - 5a) J_3\\
           & & \qquad\quad - \ 2 a (1 - a) L_8
                - 2a (1 - a) L_7 -  2 (4 + 3a) L_6 + 2 (4 - 3a) L_5\\ 
           & & \qquad\quad + \ 2a L_4  + a (10 - a) L_3
                 + (8 - 6a - a^2 ) L_2 - 2 (4 - 3a + a^2 ) L_1  \Bigr]\\
           & & + \  2 a \,f_{LL} f_{LR} \ \Bigl[ \ 2 J_{\rm IR}^1 - 8 J_3 
             + 4 L_6 + 4 L_5 + a L_3 +  a L_2 - 2 (2 - a) L_1 \Bigr] \ ,\\
  C_{0101} &=& \frac{\sqrt{a}}{2} \ (f_{LL}^2 + f_{LR}^2 ) \ \Bigl[ \
            4 J_{\rm IR}^1 - (8 + a) J_3  - (8 - a)(1 - a) L_7 - (12 + a) L_6 \\
           & & \qquad -\  2a L_5 + 2a L_4 + (8 + a) L_3 + (4 - 5a) L_2 
                  + 4 (2 - a) L_1  \Bigr]\\
           & & + \  \sqrt{a} \ f_{LL} f_{LR} \ \Bigl[ \ 4 J_{\rm IR}^1 
                     - (16 - a) J_3 - a (1 - a) L_7 + (4 + a) L_6\\
           & & \qquad\qquad\qquad + \ 2a L_5
                      + a L_3  - (4 - 3a) L_2 + 2a L_1 \Bigr]\ ,\\
  C_{1101} &=& \frac{\sqrt{a}}{2} \ C_{1000} + \sqrt{a} \ (f_{LL}^2 - f_{LR}^2 ) \\ 
           & & \times \ \Biggl[ \ 
              - a N_{10} - 9 a N_9 + 2 N_7 - 2 N_6 - \frac{1}{2} (12 + a) N_3
              + \frac{a}{2} N_2 - (12 + a) N_1\\ 
           & & \quad\quad - \ a (1 + a ) K_8 + a (1 - a) K_7
                   + 9 a (1 - a) K_6 - 3 a (5 + a) K_5\\
           & & \quad\quad - \  (12 + a) K_4 +
               (4 - 5a) K_3 - 3 (4 + 5a) K_2
                          + 3 (4 + a) (1 - a) K_1 \Biggr]\ ,
\eean
where we have defined
\bean
      J_{\rm IR}^1 &=& (2 - a) J_6 - a J_5 \ ,\\
      J_{\rm IR}^2 &=& 2 (2 - a) N_{13} - a N_{12} - a N_{11} \ ,\\
      R_1 &=&  - 4 a (1 - a) L_7 - 4 (2 - a) L_6 - 8 (1 - a) L_5\\
          & & + \ a^2 L_4 +  a (2 + 3a) L_3
        - a (2 - a) L_2 + (8 - 8a + 3a^2 ) L_1 \ ,\\
      R_2 &=& - 4 L_6 - 4 L_5 - a L_3  - a L_2 + 2(2 - a) L_1 \ ,\\ 
      R_3 &=&  a^2 N_{10} + 3a (2 + a) N_9 
                + 4 a N_3 - 2 a N_2 + 8 (1 + a ) N_1 \\
          & & + \ 2 a^2 K_8 - a^2 (1 - a) K_7 
                   - 3 a (1 - a)(2 + a) K_6 + 6 a (1 + 2a) K_5\\
          & & + \ 2 (4 + 3a) K_4 + (a^2 + 6a - 8) K_3
                   + 3 a (8 + a) K_2 - 12 a (1 - a) K_1 \ . 
\eean
Note that the integrals $J_{\rm IR}^1\,,\,J_{\rm IR}^2$ 
namely $J_5\,,\,J_6\,,\,N_{11}\,,\,N_{12}$ and $N_{13}$
contain the infrared singularity. This singularity is exactly canceled
out in the sum Eq.(\ref{txsection}) by the contributions
from $D_{klmn}$.
We are now ready to discuss our numerical results.

\section{Numerical Results}

We are now in the position to give the cross section for 
polarized top quark
production for any $e^+ e^-$ collider.
Since we did not specify the spin angle $\xi$ for 
the top quark, we can predict the polarized cross section for 
any top quark spin.
The spin configurations we will display are the helicity, the beamline
and the off-diagonal bases for center of mass energies
$\sqrt{s} = 400 \, {\rm GeV}\,,\,800 \,{\rm GeV}\,,\,1500 \, {\rm GeV}$. 

Table 1 contains the values of the maximum center of mass speed, the 
running $\alpha_s$, the tree level cross section, the next to leading order 
cross section and the fractional
$\cal{O}\mit(\alpha_s)$ 
enhancement 
of the tree level cross section $(\kappa)$
for top quark pair production in $e^{-}_{L} e^{+}$ scattering. 

\vspace{0.5cm}
\begin{center}

\begin{tabular}{|c||c|c|c|}
\hline 
$\sqrt{s}$    & $400$ GeV & $800$ GeV & $1500$ GeV \\ \hline
\hline
$\beta$                  & $0.484$  & $0.899$ & $0.972$ \\ \hline
$\alpha_{s}(s)$          & $0.0980$ & $0.0910$ & $0.0854$ \\ \hline
$\sigma_{Total}$ Tree (pb)
                         & $0.8708$ & $0.3531$ & $0.1047$ \\ \hline
$\sigma_{Total}$ $\cal{O}\mit (\alpha_{s})$ (pb)
                         &  $1.113$ & $0.3734$ & $0.1084$ \\ \hline
$\kappa$                 & $0.2783$ & $0.05749$ & $0.03534$ \\ \hline
\end{tabular}
\\[0.1in]
Table \ 1: The values of $\beta$, $\alpha_s$, tree level
 and next to leading order
           cross sections and $\kappa$ \\
\hspace*{-8cm}
            for $e^{-}_{L} e^{+}$ scattering. 
\end{center}

\vspace{.3cm}
\noindent
At $\sqrt{s} = 400$ GeV the QCD corrections enhance the total
cross section by $\sim 30 \%$ compared
to the tree level results whereas at higher energies, $800$ and $1500$ GeV,
the enhancements are at the $\sim 5 \%$ level.

First we will show the numerical values for the coefficients 
($D_{klmn} + \hat{\alpha}_{s} C_{klmn}$) in Eq.(\ref{txsection}). 
Since the dominant effect of the 
$\cal{O}\mit(\alpha_s)$
corrections is
a multiplicative enhancement of the tree level result we have
chosen to write
\[
D_{klmn}+\hat{\alpha}_{s} C_{klmn} \equiv (1+\kappa)D^0_{klmn} + S_{klmn}.
\]
The $(1+\kappa)D^0_{klmn}$ terms are the multiplicative enhancement
of the tree level result whereas the $S_{klmn}$ give the 
$\cal{O}\mit(\alpha_{s})$
deviations to the spin correlations. The numerical value of these
coefficients are given in Table \ 2. 

\begin{center}
\begin{tabular}{|c||c|c|c|}
\hline 
$\sqrt{s}$                 & $400$ GeV & $800$ GeV & $1500$ GeV \\ \hline
\hline
$(1 + \kappa)D^{0}_{0000}$ & $1.511  $  & $ 1.722  $   & $ 1.671   $ \\ \hline
$(1 + \kappa)D^{0}_{2000}$ & $0.2404 $  & $ 1.241  $   & $ 1.526   $ \\ \hline
$(1 + \kappa)D^{0}_{1000}$ & $0.7809 $  & $ 2.126  $   & $ 2.406   $ \\ \hline
$(1 + \kappa)D^{0}_{0010}$ & $0.3905 $  & $ 1.063  $   & $ 1.203   $ \\ \hline
$(1 + \kappa)D^{0}_{2010}$ & $0.3905 $  & $ 1.063  $   & $ 1.203   $ \\ \hline
$(1 + \kappa)D^{0}_{1010}$ & $1.751  $  & $ 2.963  $   & $ 3.197   $ \\ \hline
$(1 + \kappa)D^{0}_{0101}$ & $1.452  $  & $ 1.100  $   & $ 0.6186  $ \\ \hline
$(1 + \kappa)D^{0}_{1101}$ & $0.3416 $  & $ 0.4650 $   & $ 0.2807  $ \\ \hline
\hline
$S_{0000}$ & $ -0.002552  $ & $  0.0005645  $ & $  0.01172   $ \\ \hline
$S_{2000}$ & $  0.007655  $ & $ -0.001682   $ & $ -0.03516   $ \\ \hline
$S_{1000}$ & $  0.02154   $ & $  0.01224    $ & $ -0.02085   $ \\ \hline
$S_{0010}$ & $  0.01094   $ & $  0.01586    $ & $  0.008357  $ \\ \hline
$S_{2010}$ & $  0.01059   $ & $ -0.006158   $ & $ -0.03614   $ \\ \hline
$S_{1010}$ & $  0.004433  $ & $ -0.02030    $ & $ -0.06134   $ \\ \hline
$S_{0101}$ & $ -0.006564  $ & $ -0.04413    $ & $ -0.04952   $ \\ \hline
$S_{1101}$ & $  0.007920  $ & $ -0.01270    $ & $ -0.02047   $ \\ \hline
\end{tabular}
\\[0.1in]
Table \ 2: The values of $(1+\kappa)D^0_{klmn}$ and $S_{klmn}$ 
           for $e^{-}_{L} e^{+}$ scattering. 
\end{center}
The ratios $ {S_{klmn}}/{(1+\kappa)D^0_{klmn}} $ are never larger than
$10 \%$ and are typically of order a few percent.
Hence the $\cal{O}\mit(\alpha_s)$ corrections make only small
changes to the spin orientation of the top quark. 

To illustrate the different spin bases we present the top quark
production cross section in the three different spin bases
discussed in ref.~\cite{ps}.
One is the usual helicity basis which corresponds to $\cos \xi = + 1$. 
The second is the beamline basis in which the top quark spin is aligned
with the positron in the top quark rest frame
($\xi$ for the beamline basis is obtained from Eq.(\ref{offdiaxi})
with $f_{LR} ~=~ 0$).
The third corresponds to the 
off-diagonal basis which has been defined in Eq.(\ref{offdiaxi}). 
Note that as $\beta \to 1$, all of these bases coincide.
Therefore, at an extremely
high energy collider, there will be no significant difference
between these bases. 

%%%%%%%%%%%%%%%%%%%%%%%%%%%%
% Fig.8
%%%%%%%%%%%%%%%%%%%%%%%%%%%%
In Fig. 8 we give the results for $\sqrt{s} = 400$ GeV for both
$e^-_L e^+$ and $e^-_R e^+$ scattering using the helicity,
beamline and off-diagonal spin bases. This figure shows the 
tree level, SGA and the full QCD results for all
three spin bases. 
Since the results in the SGA almost coincide with the full QCD results 
the probability
that hard gluon emission flips the spin of the top quark is very small.
Clearly, the qualitative features of
the cross sections remain the same as those in the leading order
analysis.
That is the top quarks are produced with  very high
polarization in polarized $e^+e^-$ scattering.
In Table \ 3 we give the fraction of the top quarks in the sub-dominant
spin configuration for $e^-_L e^+$ scattering,
\[
\sigma 
\left(
     e^{-}_{L} e^{+} \rightarrow t_{\downarrow} X(\bar{t} {\rm ~or~} \bar{t} g) 
\right) / \sigma^{Total}_{L}
\]
for the three bases. 
Similar results also hold for $e^-_R e^+$ scattering.

\newpage
\begin{center}
\begin{tabular}{|c||c|c|c|}
\hline 
$\sqrt{s} = 400$ GeV & Helicity & Beamline & Off-Diagonal \\ \hline
\hline
Tree       & $0.336  $ & $0.0119 $ & $0.00124$ \\ \hline
SGA        & $0.332  $ & $0.0113 $ & $0.00129$ \\ \hline
$\cal{O}\mit (\alpha_{s})$       
           & $0.332  $ & $0.0115 $ & $0.00150$ \\ \hline
% dominant spin
%\hline 
% $\sqrt{s} = 400$ GeV & Helicity & Beamline & Off-Diagonal \\ \hline
%\hline
%Tree       & $0.6636$ & $0.9881$ & $0.9988$ \\ \hline
%SGA        & $0.6681$ & $0.9887$ & $0.9987$ \\ \hline
%$\cal{O}\mit (\alpha_{s})$       
%           & $0.6681$ & $0.9885$ & $0.9985$ \\ \hline
\end{tabular}
\\[0.1in]
Table \ 3: The fraction of the $e^-_L e^+$ cross section in the
           sub-dominant spin at  \\ \hspace{1.5pc}
           $\sqrt{s} = 400$ GeV for the helicity, beamline 
           and off-diagonal bases.\\ \hspace{5pc}
           For the soft gluon approximation (SGA) 
           $\omega_{\rm max}=(\sqrt{s}-2m)/5=~10$ GeV
\footnote{In the SGA 
the fractions in Tables 3, 4 and 5 have a very small
dependence on $\omega_{\rm max}$.}.
\end{center}

%%%%%%%%%%%%%%%%%%%%%%%%%%%%
% Fig.9 and 10 
%%%%%%%%%%%%%%%%%%%%%%%%%%%%
In Fig. 9 and 10 we have plotted the similar 
results for $800$ and $1500$ GeV colliders.
The fraction of top quarks in the sub-dominant spin component for 
$e^-_L e^+$ is given in Tables 4  and 5.

%
%-------------------------------------------------------------------
\begin{center}
\begin{tabular}{|c||c|c|c|}
\hline 
 $\sqrt{s} = 800$ GeV & Helicity & Beamline & Off-Diagonal \\ \hline
\hline
Tree       & $0.168  $ & $0.0690 $ & $0.0265 $ \\ \hline
SGA        & $0.164  $ & $0.0679 $ & $0.0272 $ \\ \hline
$\cal{O}\mit (\alpha_{s})$       
           & $0.165  $ & $0.0708 $ & $0.0319 $ \\ \hline
% dominant spin 
%\hline
%Tree       & $0.8318$ &  $0.9310$ & $0.9735$ \\ \hline
%SGA        & $0.8358$ &  $0.9321$ & $0.9728$ \\ \hline
%$\cal{O}\mit (\alpha_{s})$       
%           & $0.8350$ & $0.9292$ & $0.9682$ \\ \hline
\end{tabular}
\\[0.1in]
Table \ 4: The fraction of the $e^-_L e^+$ cross section in the
sub-dominant spin at  \\ \hspace{1.5pc} 
$\sqrt{s} = 800$ GeV for the helicity, beamline and off-diagonal
bases.\\ \hspace{5pc} 
For the soft gluon approximation (SGA) 
$\omega_{\rm max}=(\sqrt{s}-2m)/5=~90$ GeV.
\\[0.3in]
%-------------------------------------------------------------------
\begin{tabular}{|c||c|c|c|}
\hline 
 $\sqrt{s} = 1500$ GeV & Helicity & Beamline & Off-Diagonal \\ \hline
\hline
Tree       & $0.132  $ & $0.0978 $ & $0.0466 $ \\ \hline
SGA        & $0.130  $ & $0.0973 $ & $0.0472 $ \\ \hline
$\cal{O}\mit (\alpha_{s})$       
           & $0.133  $ & $0.101  $ & $0.0552 $ \\ \hline
%\hline
%Tree       & $0.8680$ & $0.9022$ & $0.9535$ \\ \hline
%SGA        & $0.8696$ & $0.9027$ & $0.9528$ \\ \hline
%$\cal{O}\mit (\alpha_{s})$       
%           & $0.8671$ & $0.8985$ & $0.9448$ \\ \hline
\end{tabular}
\\[0.1in]
Table \ 5: The fraction of the $e^-_L e^+$ cross section in the
sub-dominant spin at \\ \hspace{1.5pc}  
$\sqrt{s} = 1500$ GeV for the helicity, beamline and off-diagonal
bases.\\ \hspace{5pc}
For the soft gluon approximation (SGA) 
$\omega_{\rm max}=(\sqrt{s}-2m)/5=~230$ GeV.
\end{center}
%-------------------------------------------------------------------

Our numerical studies demonstrate that the QCD corrections have a small
effect on the spin configuration of the produced top (or anti-top) quark
for any spin basis.
The off-diagonal and beamline bases are clearly more sensitive to
the radiative corrections than the helicity basis. 
However, in the off-diagonal basis,
the top (and/or anti-top) quarks are produced in an essentially unique
spin configuration even after including the lowest order QCD corrections.

\section{Conclusion}

We have studied the ${\cal O} (\alpha_s )$ QCD 
corrections to top quark production in a generic spin basis. 
The QCD corrections introduce two effects
not included in the tree level approximation. 
One is the modification of the coupling of the top and anti-top quarks
to $\gamma$ and $Z$ bosons due to the virtual corrections. 
The another is the presence of the real gluon emission process. 
First, we consider the QCD corrections in the soft gluon
approximation to see the effects of the modified 
$\gamma / Z - t - \bar{t}$ vertex.
Using this approximation, we find the tree level 
off-diagonal basis continues to make the like spin components vanish 
and that the effects of the included anomalous magnetic moment are small.

Next we analyzed the full QCD corrections at one loop level. 
When we consider the three particle final state, 
the top and anti-top quarks are not necessarily
produced back to back. 
So we have calculated the inclusive top (anti-top) quark production.  
In this paper we have given an exact analytic form for the
differential cross section with an arbitrary
orientation of the top quark spin.

Our numerical studies show that
the ${\cal O} (\alpha_s )$ QCD corrections enhance the 
tree level result and only slightly modifies the spin
orientation of the produced top quark.
In the kinematical region where the emitted gluon has small
energy, it is natural to expect that the real gluon emission effects
introduce only a multiplicative correction to the tree level result.
Therefore only \lq\lq hard\rq\rq\ gluon emission could possibly modify
the top quark spin orientation. 
What we have found, by explicit calculation, is that this effect
is numerically very small. 
The size of the QCD corrections to the total cross section and the
enhancement of the tree level results
can be read off from the values of $\kappa$ in Table \ 1 of Sec.4.
At $\sqrt{s} = 400 \, {\rm GeV}$, the enhancement is $\sim 30 \%$ whereas
at higher energies, 800 and 1500 GeV, it is at the $\sim 5 \%$ level.
Near the threshold, the QCD corrections have a singular
behavior in $\beta$, the speed of the produced quark, 
this factor enhances the value of the correction at smaller energy.
The size of these corrections is reasonable for QCD,
on the other hand, the change of the orientation of top quark spin are
quite small. 
The deviation from the enhanced tree level result is less than a few percent.
We can, therefore, conclude 
that the results of the tree level analysis are not changed even
after including QCD radiative corrections except for a multiplicative
enhancement. This means that for the beamline and off-diagonal bases,
the top (and/or anti-top) quarks are produced in an essentially unique
spin configuration. 
Actually, the fraction of the top
quarks in the dominant (up) spin configuration for $e_L^- e$
scattering is more than $94 \%$ at all energies we have considered. 

As has been discussed in many articles,
there are strong correlations between the orientation of the
spin of the produced top (anti-top) quark and the angular distribution
of its decay products.
Therefore, measuring the top quark spin orientation will give us 
important information on the top quark sector of the Standard Model
as well as possible physics beyond the Standard Model.

\section{Acknowledgments}
\noindent
The Fermi National Accelerator Laboratory is operated by the
University Research Association, Inc., under contract
DE-AC02-76CHO3000 with the United States of America Department of
Energy. J.K. was supported in part by the Monbusho Grant-in-Aid for
Scientific Research (Japan) No.C(2) 09640364. J.K. and T.N. would like 
to thank W. Bardeen for the hospitality extended to them at FNAL where 
part of this work was performed.
S.P. would like to thank J. Kodaira for the hospitality extended to him
while at Hiroshima University.

\newpage
\section{Appendix A : Phase space integrals over $y$ and $z$}

\noindent
The phase space integrals necessary to derive the cross section
are summarized in this Appendix. Although many of them have already appeared
in the literatures~\cite{s,tung}, we will list all of them below
for the convenience of the reader. After the integration over the angular
variables, we are left with the following four types of integrals:
\bean
   J_i &=& \int dy dz f_i (y , z)\ , \qquad
     N_i = \int \frac{dy dz}{\sqrt{(1 - y)^2 - a}} f_i (y , z)\ ,\\
   L_i &=& \int \frac{dy dz}{(1 - y)^2 - a} f_i (y , z) \ , \quad
     K_i = \int \frac{dy dz}{\{ (1 - y)^2 - a \}^{3/2}} f_i (y , z) \ .
\eean
The infrared divergences are regularized by the small gluon mass
$\lambda$ and $\beta = \sqrt{1 - a} = \sqrt{1 - 4 m^2 / s}$.
For the type $L_i$ integrals, a shorthand notation
$\omega = \sqrt{ (1 - \sqrt{ a})/(1 + \sqrt{ a}) }$ is used.
The $K_i$ integrals have a spurious singularity at the upper bound of
the $y$ integral, $y_+ = 1 - \sqrt{a}$. Since this singularity turns out
to be canceled out in the cross section, we regularize each integrals
by deforming the integration region
as $y_+ \to 1 - \sqrt{a} - \epsilon$~\cite{s}. 
${\rm Li}_2$ is the Spence function.

\vspace{1cm}
\noindent
Class {\boldmath{$J$}} Integrals:
\bean
      \J{1} & = & 
      \uint \ = \ 
      \frac{1}{2} \beta \left(1+\frac{1}{2} a\right)-\frac{1}{2} a\left(
      1-\frac{1}{4} a\right)\ln\left({1+ \beta \over 1- \beta }\right)\\
      \J{2} & = &
      \uint {y\over z}\ =\ \uint {z\over y}\\
      & = &
      -\frac{1}{4} \beta \left(5-\frac{1}{2} a\right)+\frac{1}{2}\left(
      1+\frac{1}{8} a^2\right)\ln\left({1+ \beta \over 1- \beta }\right)\\
      \J{3} & = &
      \uint {1\over y}\ =\ \uint {1\over z} = 
      - \beta +\left(1-\frac{1}{2} a\right)\ln\left({1+ \beta \over 1- 
        \beta }\right) \\
     \J{4} & = & \uint \frac{y}{z^{2}} = \uint \frac{z}{y^{2}} =
     \frac{2}{ a} \beta  - \ln \left(\frac{1+ \beta }{1- \beta } \right) \\
      \J{5} & = &
      \uint {1\over y^2}\ =\ \uint {1\over z^2} \\
      & = &
      -{2 \beta \over a}\left(
      \ln \frac{\lambda^2}{s} + 2 \ln a-4\ln \beta -4\ln 2+2\right)+
      2\left(1-{3\over a}\right)\ln\left({1+ \beta \over 1-\beta }\right)\\
      \J{6} & = &
      \uint {1\over y z} \\
      & = &
      \left(- \ln \frac{\lambda^2}{s} -\ln a+4\ln \beta +2\ln 2\right)
      \ln\left({1+ \beta \over 1- \beta }\right)  \\
      & &
     +2\left[\Li\left({1+ \beta \over 2}\right)
      -\Li\left({1- \beta \over 2}\right)\right]
     +3\left[\Li\left(-{2 \beta \over 1- \beta }\right)
       -\Li\left({2 \beta \over 1+ \beta }\right)\right]
\eean

\vspace{1cm}
\noindent
Class {\boldmath{$N$}} Integrals:
\bean
      \N{1} & = & \pint  = 
      1-\sqrt{ a}-\frac{1}{2} a\ln\left({2-\sqrt{ a}\over\sqrt{ a}}\right)\\
      \N{2} & = & \pint {z\over y} =
      -\frac{1}{2}\ln a+\ln\left(2-\sqrt{ a}\right)+
      {2\over 2-\sqrt{ a}}-2 \\
      \N{3} & = &
      \pint {y\over z}\\
      & = &
      2 \beta \ln\left({1- \beta \over 1+ \beta }\right)
      -\ln\left({1+ \beta \over 2}\right)\ln\left({1- \beta \over 2}\right)
      +\left(2-\frac{a}{2}\right)
      \ln\left({2-\sqrt{ a}\over\sqrt{ a}}\right)\\
      & & +\, \frac{1}{4} \ln^2 \frac{a}{4} -\sqrt{ a} + 1
      + \Li\left({1+ \beta \over 2}\right)+\Li\left({1- \beta \over 2}\right)-
      2\,\Li\left({\sqrt{ a}\over 2}\right)\\ 
      \N{4} & = & \pint {y^2\over z^2}  = 
      {2\over a} \left(1-\sqrt{ a}\right)^2 \\
      \N{5}& = & \pint y \\
      & = &
      {1\over 16} \left[ - a^2\ln a+2\, a^2\ln(2-\sqrt{ a})+
      4(2-\sqrt{ a})^2-4\left(2- a^{3\over2}\right)\right]\\
      \N{6} & = & \pint z \\
      & = &
      {1\over 32} \left[ 12-(2+ a)^2-{2+\sqrt{ a}\over 2-\sqrt{ a}}\, a^2
      +2(8- a) a\,\ln{\sqrt{ a}\over 2-\sqrt{ a}} \right] \\
      \N{7} & = & \pint {y^2\over z} \\
      & = &
      \left(1+\frac{1}{2} a\right) \left[ \Li\left({1+ \beta \over 2}\right)+
      \Li\left({1- \beta \over 2}\right)-2\,\Li\left(\frac{1}{2}\sqrt{ a}\right)+
      \frac{1}{4}\ln^2\left(\frac{1}{4} a\right) \right. \\
      & &
      \left. -\ln\left({1+ \beta \over 2}\right)
         \ln\left({1- \beta \over 2}\right) \right]+
      3v \ln\left({1- \beta \over 1+ \beta }\right)+ \frac{1}{8}(18+ a) \\
      & &
      -\frac{1}{8}(20- a)\sqrt{ a} + \left(3- a+\frac{1}{16}
       a^2\right) \ln\left({2-\sqrt{ a}\over\sqrt{ a}}\right) \\
      \N{8} & = & \pint {1\over y}  =
      2 \ln\left({2-\sqrt{ a}\over\sqrt{ a}}\right) \\
      \N{9} & = &
      \pint {1\over z} \\
      & = &
      \Li\left({1+ \beta \over 2}\right)+\Li\left({1- \beta \over 2}\right)+
      2\,\Li\left(-{\sqrt{ a}\over 2-\sqrt{ a}}\right)+\frac{1}{4} \ln^2 a  \\
      & &
      +\ln^2\left({2-\sqrt{ a}\over 2}\right)-\ln(1+ \beta )\ln(1- \beta ) \\
      \N{10} & = & \pint {y\over z^2}  = 
      {4\over a} \left(1-\sqrt{ a}\right) \\
      \N{11} & = & \pint {1\over y^2} \\
      & = &
      {2\over a} \left[ -\ln \frac{\lambda^2}{s} + \ln a +
      2 \ln(1-\sqrt{ a})- 4\ln(2-\sqrt{ a})+ 2 \ln 2- 2 \right] \\
      \N{12} & = & \pint {1\over z^2} \\
      & = &
      {2\over a} \left[ -\ln \frac{\lambda^2}{s} - \ln a +
      2 \ln(1-\sqrt{ a})-{1+ \beta^2 \over \beta}
      \ln\left({1+ \beta \over 1- \beta }\right)+ 2 \ln 2 \right] \\
      \N{13} & = & \pint {1\over y z} \\
      & = &
      {1\over \beta}\ln\left({1- \beta \over 1+ \beta}\right)
       \left[ \ln \frac{\lambda^2}{s} +
      \frac{1}{2}\ln a+4\ln(2-\sqrt{ a})-4\ln (2 \beta ) -
      2\ln\left({1- \beta \over 1+ \beta}\right) \right] \\
      & &
      + \, {1\over \beta}\ln^2\left({(1-\beta )^2\over\sqrt{ a}
         (2-\sqrt{ a})} \right) +
      {2\over \beta } \ln\left({\sqrt{ a}(2-\sqrt{ a})\over 2}\right)
      \ln\left({2\sqrt{ a}(1-\sqrt{ a})\over (1-\sqrt{ a}- \beta )^2}\right) \\
      & &
      + \, {2\over \beta} \left[ 
        \Li\left({\sqrt{ a}(2-\sqrt{ a})\over (1+ \beta )^2}
      \right)-\Li\left[\left({1- \beta \over 1+ \beta }\right)^2\right]+
      \Li\left({(1- \beta )^2\over\sqrt{ a}(2-\sqrt{ a})}\right) \right] \\
      & &
      + \, {1\over \beta } \left[ \Li\left({1+ \beta \over 2}\right)
      + \Li\left(-{2 \beta \over 1- \beta }\right) - (\beta \to -
            \beta ) - {\pi^2\over 3} \right]
\eean

\vspace{1cm}
\noindent
Class {\boldmath{$L$}} Integrals:
\bean
      \L{1} & = & \tint =
      2\,{1 - a \over 4 - a}\ln\left({1+ \beta \over 1- \beta }\right) \\
      \L{2} & = & \tint  {z\over y} \\
      & = &
      \left(-1+{12\over 4- a}-{24\over(4- a)^2}\right)
      \ln\left({1+ \beta \over 1- \beta }\right)
      -\frac{2 \beta }{4- a} \\
      \L{3} & = &
      \tint {y\over z} \\
      & = &
      \frac{1}{2}\ln\left(\frac{1+ \beta }{1- \beta }\right)
      \Biggl[\,\frac{1}{2}\ln a+\ln(2+\sqrt{a})-\ln(1+\sqrt{a})-2\ln2\,\Biggr]
      \nonumber \\ &&
      + \, {1-\sqrt{a} \over\sqrt{a} }\Biggl[\,\Li(\omega )
      + \Li\left({2+\sqrt{a} \over 2-\sqrt{a}}\;\omega \right)
      - ( \omega \to - \omega )\,\Biggr] \\
      & &
      + \, \Biggl[ \, \Li\left({1+\omega \over 2}\right)
      + \Li\left((2+\sqrt{a} ){1+\omega \over 4}\right)
      + \Li\left({2\sqrt{a} \over (2+\sqrt{a} )(1+\omega )}\right)\\
      & &  \qquad\qquad\qquad\qquad - \, ( \omega \to - \omega ) \, \Biggr] \\
      \L{4} & = &  \tint {y^2\over z^2}  =
      \frac{2}{a}\left[\,\ln\left({1+ \beta \over 1-\beta }\right)
          -2 \beta \,\right] \\
      \L{5} & = & \tint y  = 
      -\left(1+\frac{1}{2} a-{6\over 4- a}\right)
      \ln\left({1+ \beta \over 1-\beta }\right)- \beta \\
      \L{6} & = & \tint z  =
      -{3 a\over 4- a}\left(1-{2\over 4- a}\right)
      \ln\left({1+ \beta \over 1-\beta }\right)+
      {2 \beta \over 4- a} \\
      \L{7} & = & \tint {\:1\over z} 
       = \frac{1}{\sqrt{a}} \, \Biggl[\,\Li(\omega )
      +\Li\left({2+\sqrt{a} \over 2-\sqrt{a} }\;\omega \right)
      - (\omega \to - \omega ) \,\Biggr] \\
      \L{8} & = & \tint {y\over z^2} =
      \frac2 a\ln\left({1+ \beta \over 1-\beta }\right)
\eean

\vspace{1cm}
\noindent
Class {\boldmath{$K$}} Integrals:
\bean
     \K{1} & = & \kint \\
     &=& \frac{2 (1-\sqrt{ a})}{\sqrt{ a}(2-\sqrt{ a})^2}
     \ln \frac{2 \sqrt{ a}(1-\sqrt{ a})}{\epsilon (1+\sqrt{ a})}-
     \frac{4 a}{(4- a)^2}
     \ln \frac{(1+\sqrt{ a })(2-\sqrt{ a })^2}{2 \sqrt{ a } a} \\
     \K{2} & = & \kint y \\
     &=& \frac{2 (1-\sqrt{ a})^2}{\sqrt{ a}(2-\sqrt{ a})^2}
     \ln \frac{2 \sqrt{ a}(1-\sqrt{ a})}{\epsilon (1+\sqrt{ a})}+
     \frac{  a ^2}{(4-  a)^2}
     \ln \frac{(1+\sqrt{ a})(2-\sqrt{ a})^2}{2\sqrt{ a} a)}\\
     & & +\, \ln\frac{2 \sqrt{ a }}{(1+\sqrt{ a })} \\
     \K{3} & = & \kint z \\
     &=& \frac{2 (1-\sqrt{ a})^2}{(2-\sqrt{ a})^3}
     \ln \frac{2 \sqrt{ a}(1-\sqrt{ a})}{\epsilon (1+\sqrt{ a})}+
     \frac{ a( a^2+20 a-32)}{2(4- a)^3}
     \ln \frac{(1+\sqrt{ a})(2-\sqrt{ a})^2}{2\sqrt{ a} a} \\
     && - \, \frac{1}{2} \ln\frac{2 \sqrt{ a}}{(1+\sqrt{ a})}+
     \frac{2 a(1-\sqrt{ a})}{(4- a)(2-\sqrt{ a})^2}  \\
     \K{4} & = & \kint yz \\
     &=& \frac{2 (1-\sqrt{ a})^3}{(2-\sqrt{ a})^3}
     \ln \frac{2 \sqrt{ a}(1-\sqrt{ a})}{\epsilon (1+\sqrt{ a})}+
     \frac{ a^2(12-7 a)}{2(4- a)^3}
     \ln \frac{(1+\sqrt{ a})(2-\sqrt{ a})^2}{2\sqrt{ a} a} \\
     && -\,  \frac{1}{2} \ln\frac{2 \sqrt{ a}}{(1+\sqrt{ a})}+
     \frac{2}{(4- a)} 
     \Biggl[  a \frac{(1-\sqrt{ a})^2}{(2-\sqrt{ a})^2}+
     \sqrt{ a}-1 \Biggr] \\
     \K{5} & = & \kint \frac{y}{z} \\
     &=&  \frac{2 (1-\sqrt{ a})}{ a(2-\sqrt{ a})}
     \ln \frac{2 \sqrt{ a}(1-\sqrt{ a})}{\epsilon (1+\sqrt{ a})}+
     \frac{(4-3 a)}{ a(4- a)}
     \ln \frac{(1+\sqrt{ a})(2-\sqrt{ a})^2}{2\sqrt{ a} a} \\
     && -\, \frac{1}{ a} \ln\frac{2 \sqrt{ a}}{(1+\sqrt{ a})}+
     \frac{4(1-\sqrt{ a})}{ a(2-\sqrt{ a})}-
     \frac{2\beta }{ a} \ln \left( \frac{1+\beta }{1-\beta } \right) \\
     \K{6} & = & \kint \frac{1}{z}\\
     &=& \frac{2}{ a(2-\sqrt{ a})}
     \ln \frac{2 \sqrt{ a}(1-\sqrt{ a})}{\epsilon (1+\sqrt{ a})}+
     \frac{(4+ a)}{ a(4- a)}
     \ln \frac{(1+\sqrt{ a})(2-\sqrt{ a})^2}{2 \sqrt{ a} a} \\
     && - \, \frac{1}{ a} \ln\frac{2 \sqrt{ a}}{(1+\sqrt{ a})}+
     \frac{4}{ a(2-\sqrt{ a})}-
     \frac{2}{ a \beta } \ln \left( \frac{1+ \beta }{1- \beta } \right) \\
     \K{7} & = & \kint \frac{y}{z^2} =
     \frac{2}{ a \sqrt{ a}} \ln\frac{2 \sqrt{ a}(1-\sqrt{ a})}
            {\epsilon (1+\sqrt{ a})} \\
     \K{8} & = & \kint \frac{y^2}{z^2} \\
     &=&  \frac{2(1 - \sqrt{ a})}{ a\sqrt{ a}}
     \ln \left(\frac{2 \sqrt{ a}(1-\sqrt{ a})}
                    {\epsilon (1+\sqrt{ a})}\right) +
     \frac{4}{  a } \ln \left(\frac{2 \sqrt{ a}}{1+\sqrt{ a}}\right)
\eean
%
%%%%%%%%%%%%%%%%%%%%%%%%%%%%%%%%%%%%%%%%%%%%%%%
\newpage
%
%\noindent
%{\Large\bf Appendix B : Total cross section}

\section{Appendix B : Total cross section}

\noindent
Let us examine the unpolarized total
cross section including the top quark pairs using our formulae.
It is given by
\[  \sigma_T (e^- e^+ \to t + \bar{t} + X)
      = \frac{1}{4} \left[ \sigma_T (e^-_L e^+_R \to t + \bar{t} +X)
               + \sigma_T (e^-_R e^+_L \to t + \bar{t} + X) \right] \ .\]
Note that only $k=0, 2$ and $l,m,n = 0$ terms in Eq.(\ref{txsection})
contribute to the total cross section.
Integrating over the angle $\theta$, we get
\bean
  \lefteqn{\sigma_T (e^-_L e^+_R \to t + \bar{t} + X)}\\
    &=& \frac{\pi \alpha^2}{s} \beta \Biggl[
      \left( f_{LL} + f_{LR} \right)^2 \left( 3 - \beta^2 \right)\\
    & & \qquad \times \ 
         \left( 1 + \hat{\alpha}_s \left\{ V_I 
          - \frac{6}{3 - \beta^2} V_{II} + \frac{2}{\beta} J_{IR}^1
          -  \frac{8}{\beta} J_3 + \frac{8}{\beta (3 - \beta^2 )} J_2
            \right\} \right)\\
      & & + \  2 \left( f_{LL} - f_{LR} \right)^2 \beta^2\\
      & & \qquad \times \ 
           \left( 1 + \hat{\alpha}_s 
        \left\{ V_I + 2 V_{II} + \frac{2}{\beta} J_{IR}^1 -
             \frac{8}{\beta} J_3 
          + \frac{2 (3 - \beta^2 )}{\beta^3} J_2 + \frac{2 a}
                    {\beta^3} J_1 \right\} \right) \Biggr].
\eean
The cross section for the process $e^-_R e^+_L$ is obtained by
interchanging the coupling constant $L \leftrightarrow R$ in the above
expression.
Parameterizing the total cross section as
\bean
     R_t (s) &\equiv& \left( \frac{1}{\sigma_{\rm pt}} \right)
             \sigma_T (e^- e^+ \to t + \bar{t} + X)\\
            &=& \ R_t^{(0)} (s) + \frac{\alpha_s (s)}{\pi}
           C_2 (R) R_t^{(1)} (s) + \cdots \ ,
\eean
where $\sigma_{\rm pt} = 4 \pi \alpha^2 / 3s$, we get the following
numerical results at the CM energies $\sqrt{s} = 400\,,\,500\,,\,1500$ GeV.

\begin{center}
\begin{tabular}{|c||c|c|}
\hline
  \quad $\sqrt{s}$ ({\rm GeV}) \quad & \quad $R_t^{(0)} (s)$ \quad &
          \quad $C_2 (R) R_t^{(1)} (s)$ \quad\\ \hline
\hline
    400       &     1.0083      &    8.9963  \\ \hline 
    500       &     1.4190      &    6.0267  \\ \hline
   1500       &     1.7714      &    2.2911 \\ 
   \hline
\end{tabular} 
\end{center}

\noindent
These are consistent with the results in Ref.~\cite{hs}.

%%%%%%%%%%%%%%%%%%%%%%%%%%%%%%%%%%%%%%%%%%%%%%%%%

%------------------------ Figures -------------------------------
\begin{figure}[H]
\begin{center}
\begin{tabular}{cc}
\leavevmode\psfig{file=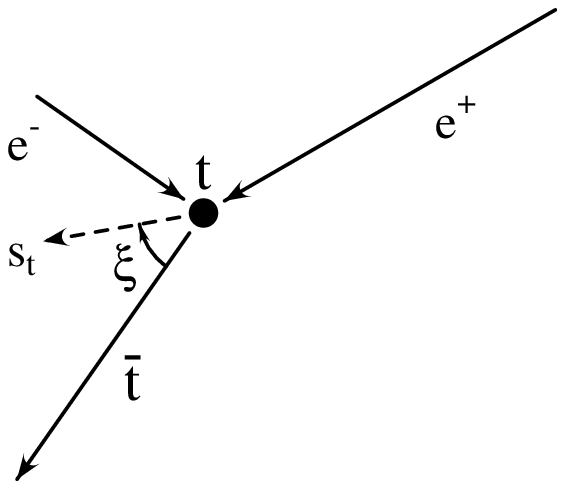,width=5.5cm} &
\leavevmode\psfig{file=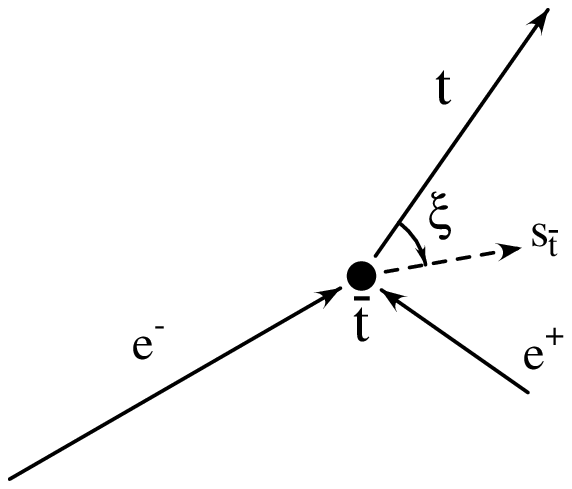,width=5.5cm} 
\end{tabular}
\caption{The generic spin basis for the top (anti-top)
quark in its rest frame.  ${\bf s}_t$ (${\bf s}_{\bar{t}}$) is the top
(anti-top) spin axis.}
\end{center}
\end{figure}

%%%%%%%%%%%%%%%%%%%%%%%
\begin{figure}[H]
\begin{center}
\begin{tabular}{cc}
\leavevmode\psfig{file=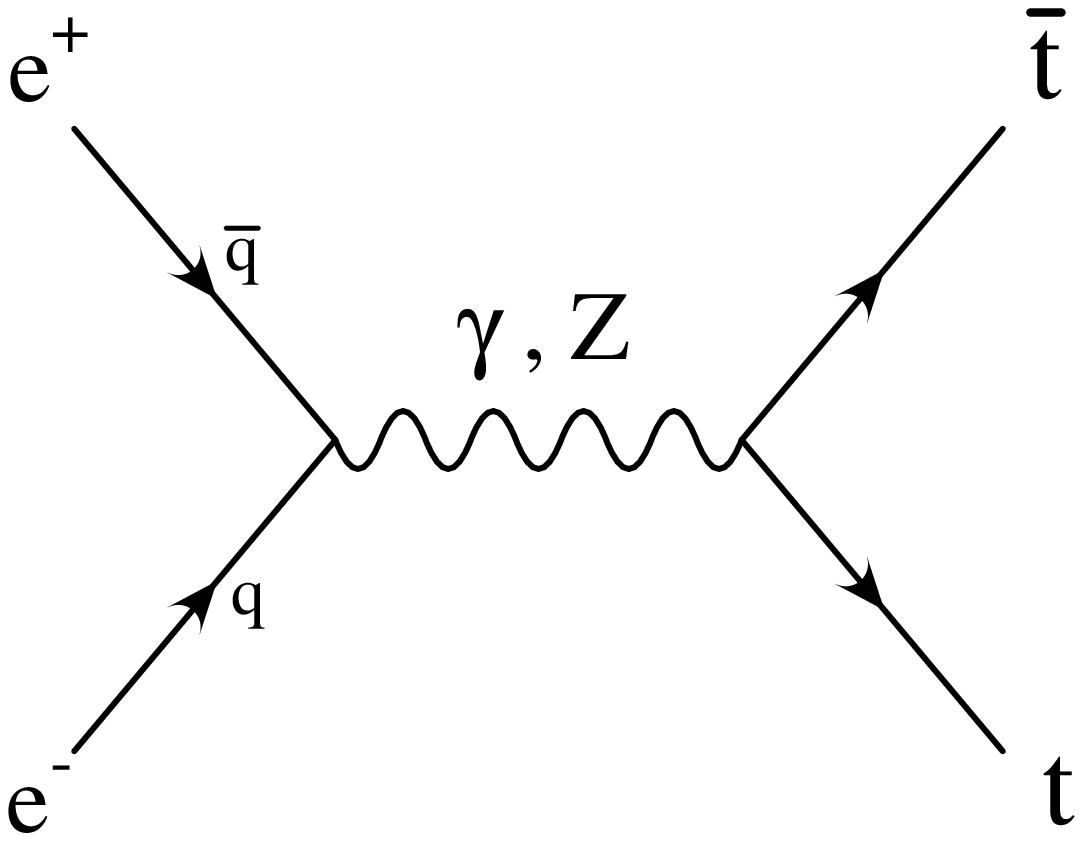,width=5.5cm} &
\leavevmode\psfig{file=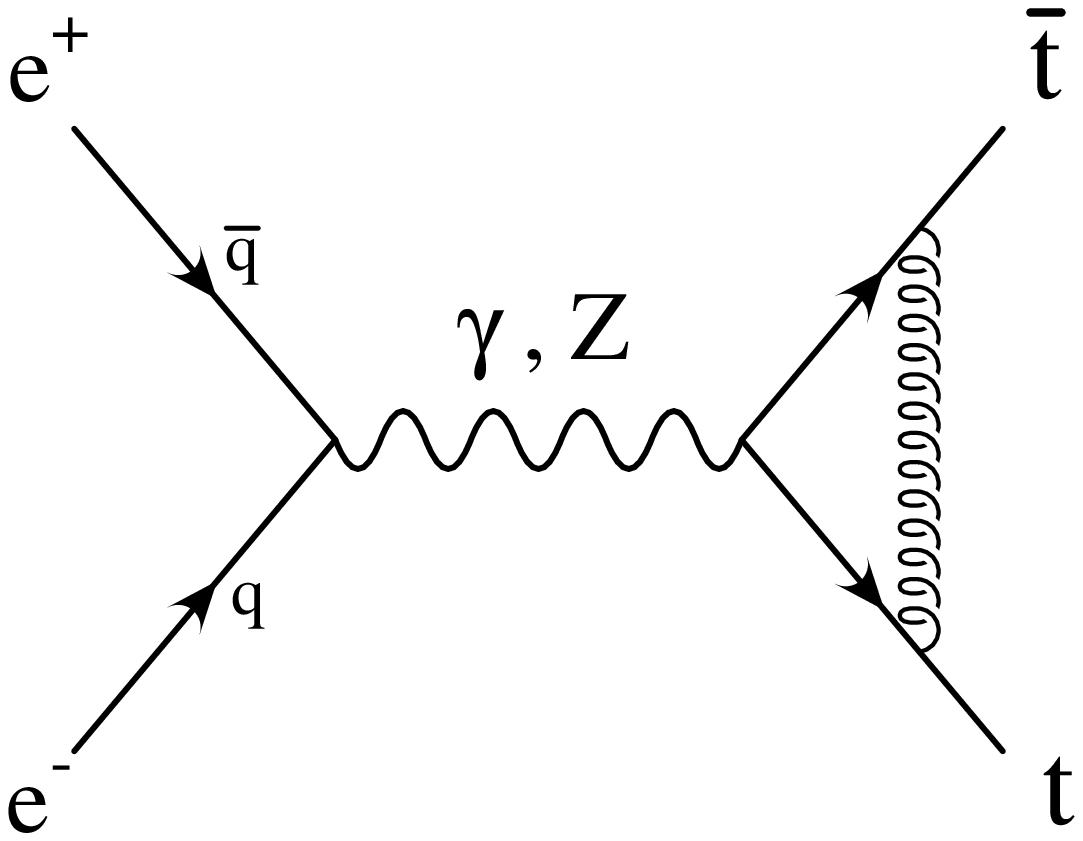,width=5.5cm} 
\end{tabular}
\caption{The tree and the QCD one-loop contributions
to the $e^- e^+ \to t \bar{t}$ process.}
\end{center}
\end{figure}
\clearpage
\begin{figure}[H]
\begin{center}
        \leavevmode\psfig{file=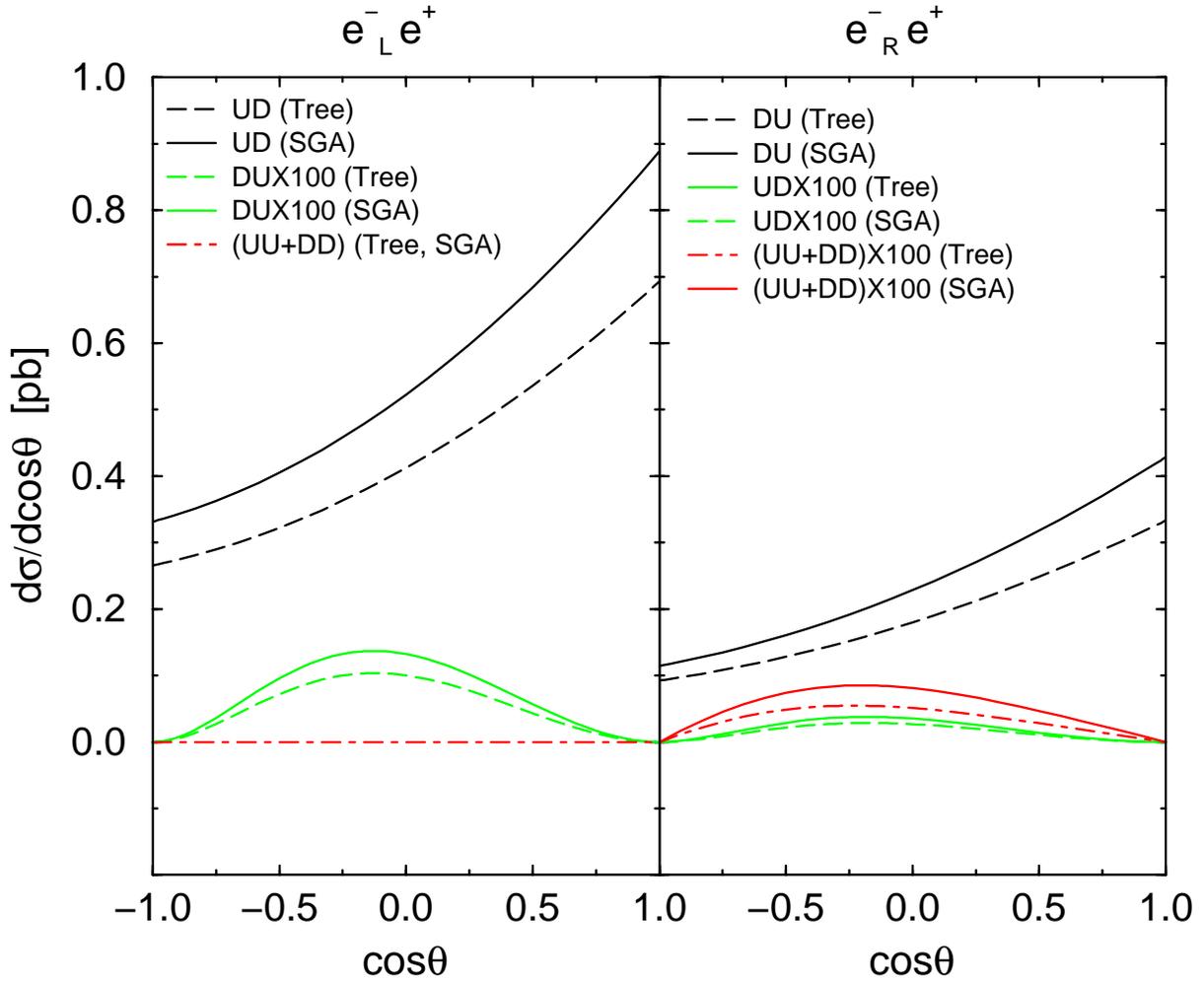,angle=-90,width=16cm}
\caption{The cross sections in the off-diagonal basis,
 Eq.(\ref{offdiaxi}), at  $\sqrt{s}=400 \, {\rm GeV}, 
\omega_{\rm max}=10 \,{\rm GeV}$ for the 
$e^{-} e^+ \rightarrow t \bar{t}$ process:
 $t_{\uparrow} \bar{t}_{\downarrow}$ (UD),
 $t_{\downarrow} \bar{t}_{\uparrow}$ (DU) and 
 $t_{\uparrow} \bar{t}_{\uparrow} 
+ t_{\downarrow} \bar{t}_{\downarrow}$ (UU$+$DD).
The suffix ``Tree'' and ``SGA'' mean the differential cross-section at the
tree level and at the one-loop level in the soft gluon
approximation. It should be noted that DU (UD) component for
the $e^{-}_{L} e^{+}$ ($e^{-}_{R} e^{+}$) process is multiplied by $100$.} 
\label{fig:fig3}
\end{center} 
\end{figure}
%%%%%%%%%%%%%%%%%%%%%%%%%%%%
\begin{figure}[H]
\begin{center}
        \leavevmode\psfig{file=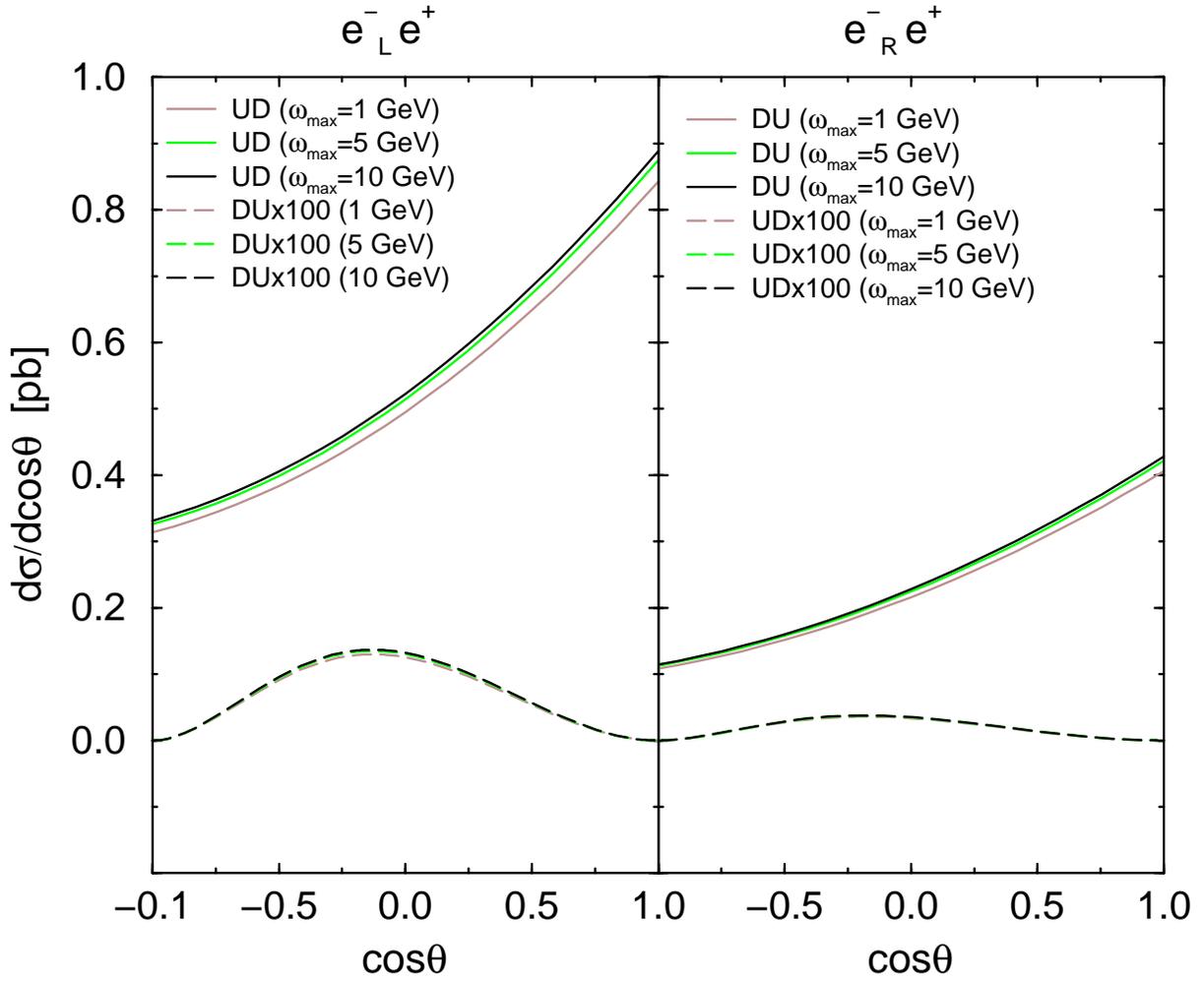,angle=-90,width=16cm}
\caption{The $\omega_{\rm max}$ dependence of the cross-sections
in the off-diagonal basis at $\sqrt{s}=400 \, {\rm GeV}$.
The DU (UD) component for
the $e^{-}_{L} e^{+}$ ($e^{-}_{R} e^{+}$) process is multiplied by $100$.} 
\label{fig:fig4}
\end{center} 
\end{figure}
%%%%%%%%%%%%%%%%%%%%%%%%%%%%%%%%%
\begin{figure}[H]
\begin{center}
\begin{tabular}{cc}
\leavevmode\psfig{file=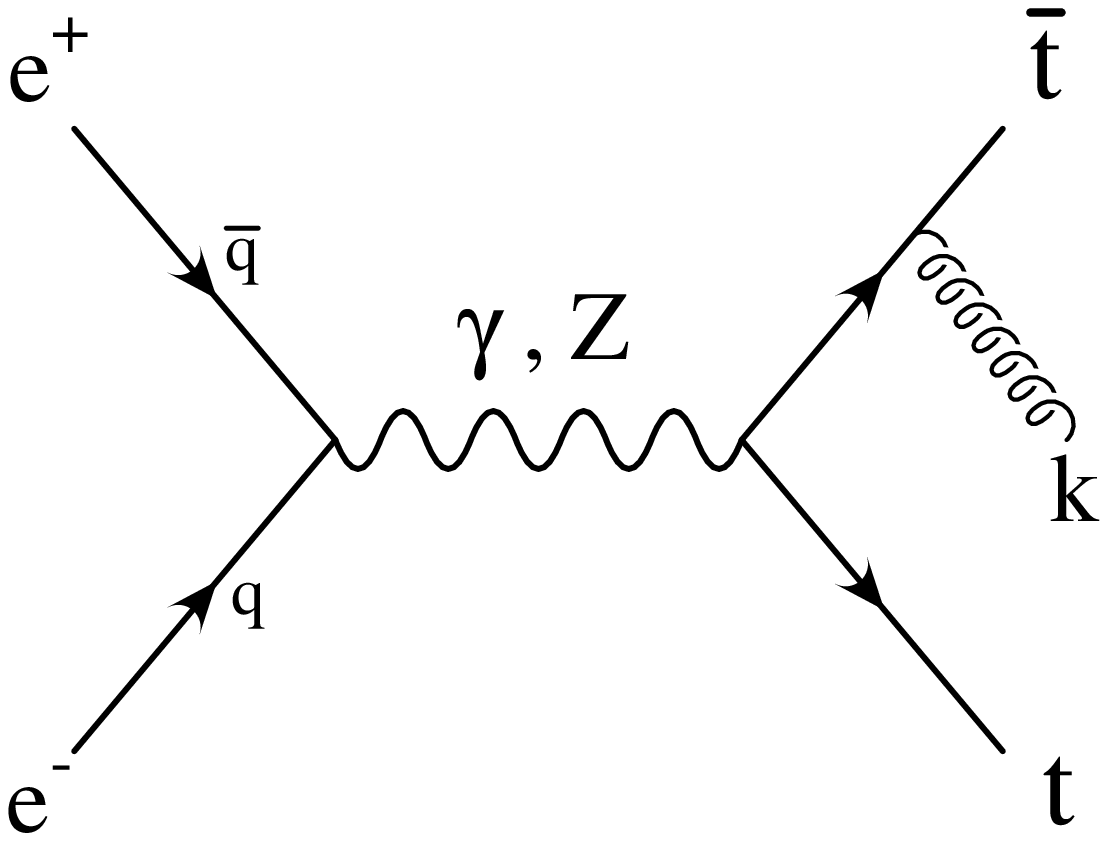,width=5.5cm} &
\leavevmode\psfig{file=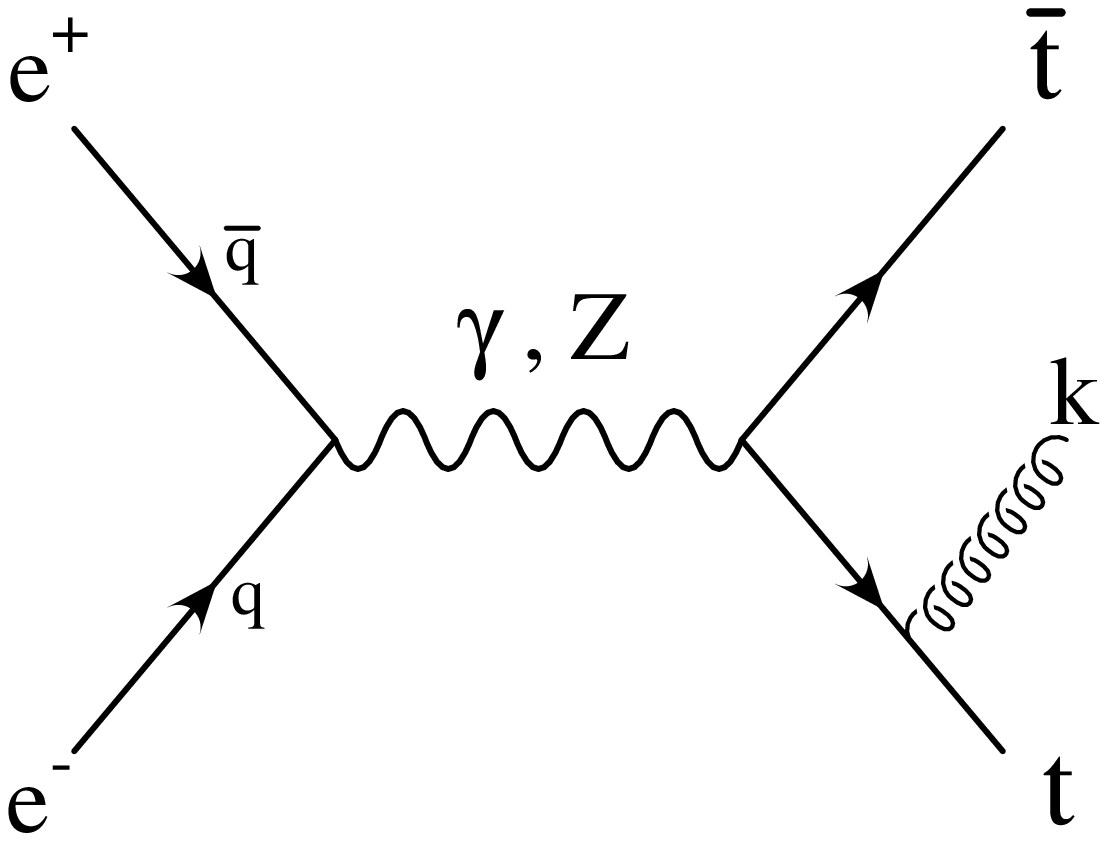,width=5.5cm} 
\end{tabular}
\caption{The real gluon emission contributions to top quark pair production.}
\end{center}
\end{figure}
%%%%%%%%%%%%%%%%%%%%%%%%%%%%%
\begin{figure}[H]
\begin{center}
\leavevmode\psfig{file=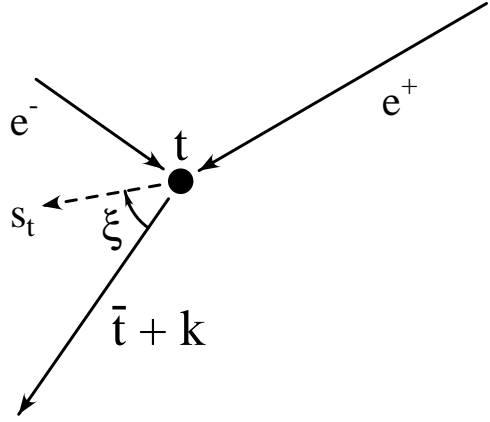,height=5.5cm}
\caption{The spin basis for the top
quark in the process  $e^- e^+ \to t \bar{t} g$.} 
\end{center}
\end{figure}
%%%%%%%%%%%%%%%%%%%%%%%%%%%%%%%%%
\begin{figure}[H]
\begin{center}
\leavevmode\psfig{file=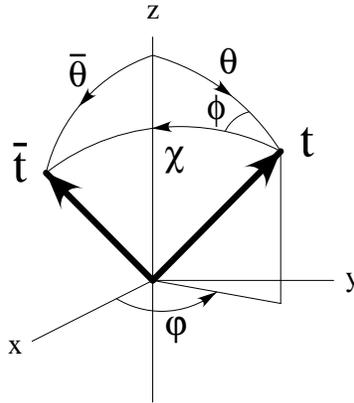,height=5.5cm}
\caption{The momentum (unit vectors) configuration of the top
and anti-top quarks in the CM frame. The momentum of $e^-$ ($e^+$)
is in the $+z$ ($-z$) direction.} 
\end{center}
\end{figure}
%%%%%%%%%%%%%%%%%%%%%%%%%%%%%%%%%
\begin{figure}[H]
\begin{center}
        \leavevmode\psfig{file=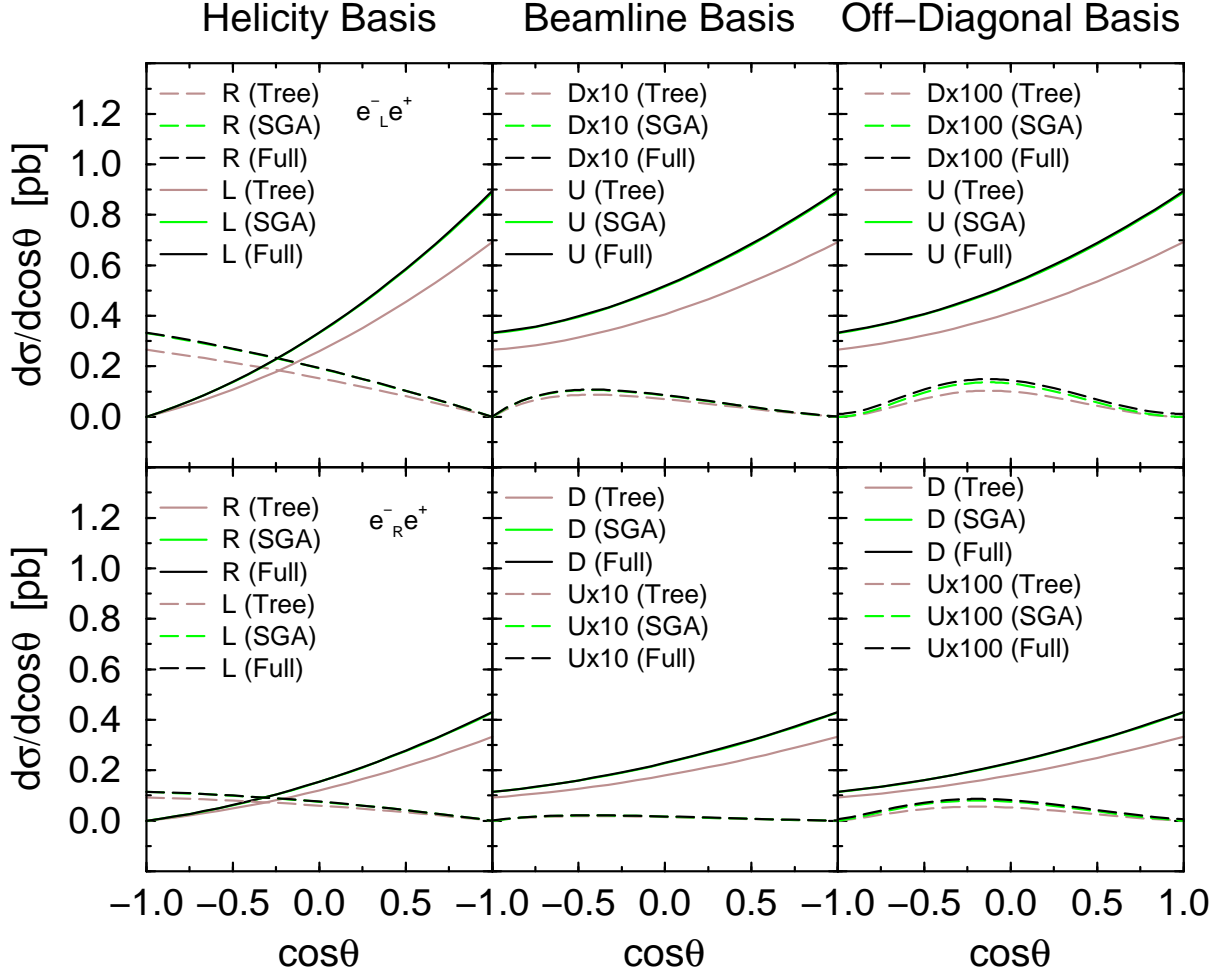,angle=-90,width=16cm}
\caption{The cross sections in the helicity, beamline and off-diagonal
 bases at $\sqrt{s} = 400 \, {\rm GeV}$.
Here we use a ``beamline basis'', in which the top quark axis is the
 positron direction in the top rest frame, for both $e^{-}_{L}e^{+}$ and 
$e^{-}_{R}e^{+}$ scattering.
For the SGA curves we have used $\omega_{\rm max}= ~10$ GeV.}
\end{center} 
\end{figure}
%%%%%%%%%%%%%%%%%%%%%%%%%%%%%%%%%
\begin{figure}[H]
\begin{center}
        \leavevmode\psfig{file=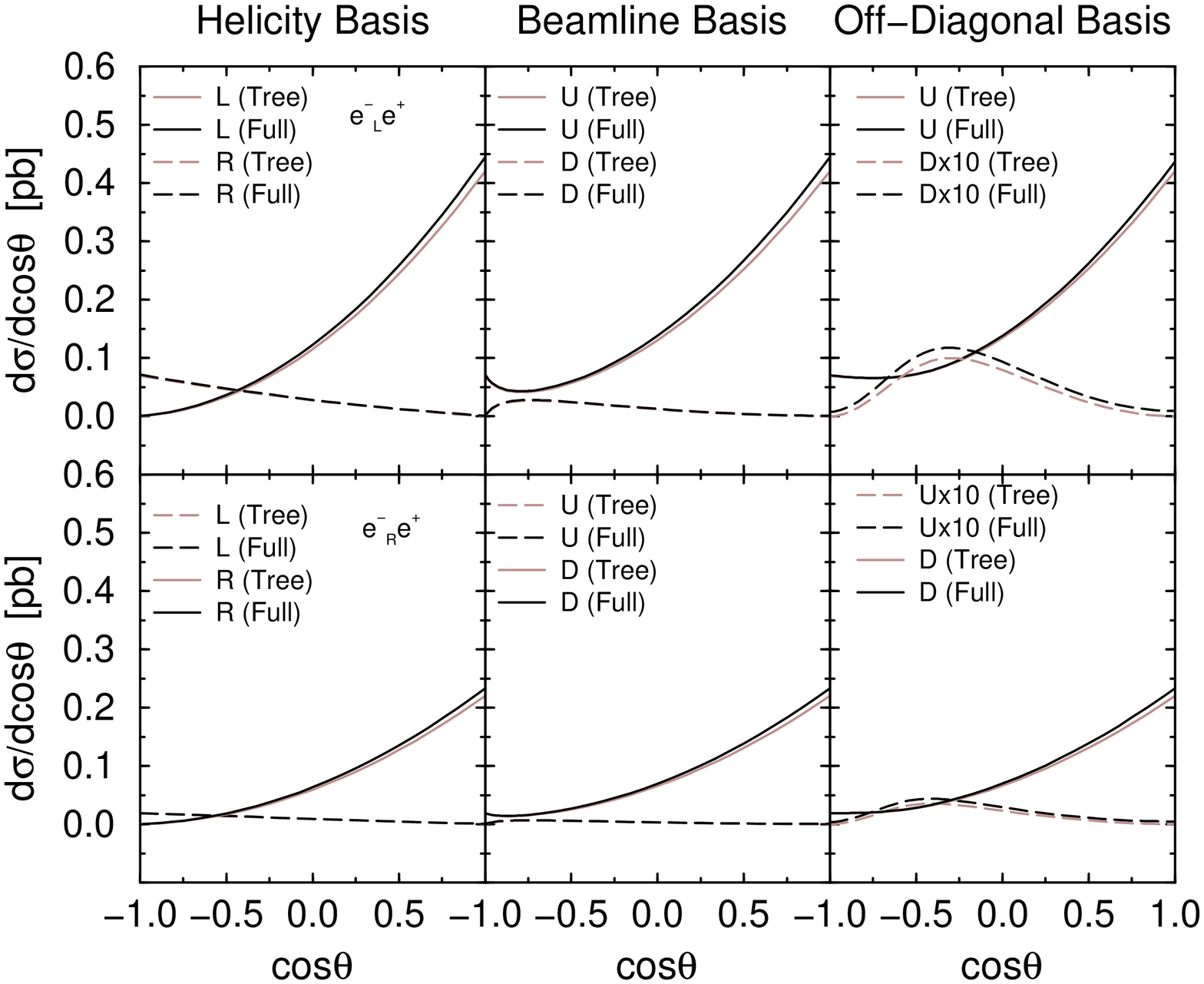,angle=-90,width=16cm}
\caption{The cross sections in the helicity, beamline and off-diagonal bases
at $\sqrt{s} = 800 \, {\rm GeV}$.}
\end{center} 
\end{figure}
%%%%%%%%%%%%%%%%%%%%%%%%%%%%
\begin{figure}[H]
\begin{center}
        \leavevmode\psfig{file=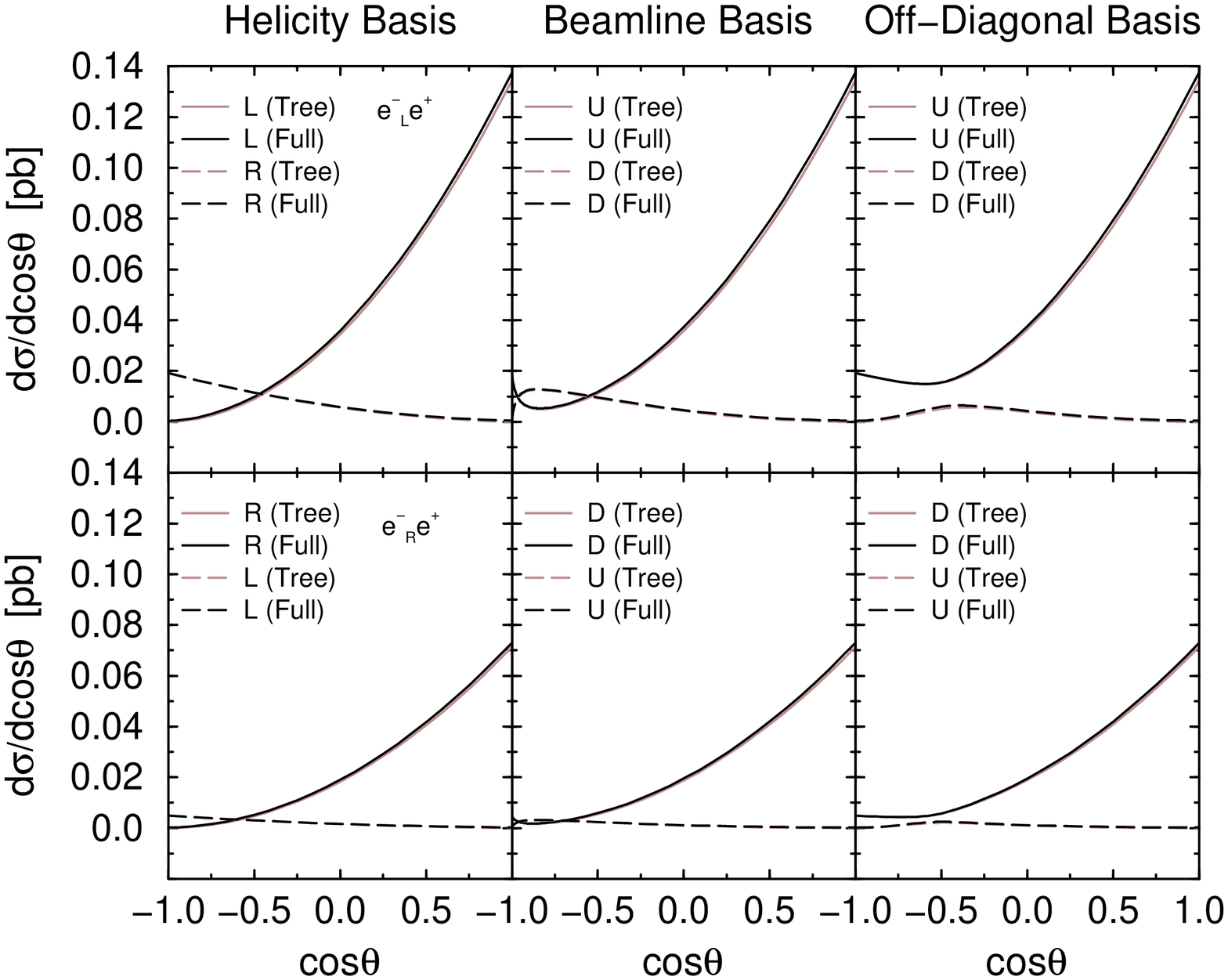,angle=-90,width=16cm}
\caption{The cross sections in the helicity, beamline and off-diagonal bases
at $\sqrt{s} = 1500 \, {\rm GeV}$.}
\end{center} 
\end{figure}
%----------------------------------------------------------------------------
\end{document}